\documentclass[a4paper,11pt]{article}
\pdfoutput=1

\usepackage{jheppub}
\usepackage[T1]{fontenc}
\usepackage[font=footnotesize,labelfont=bf]{caption}
\usepackage{graphicx}
\usepackage{subcaption}
\usepackage{bm}
\usepackage{indentfirst}
\usepackage{amsmath}
\usepackage{dirtytalk}
\usepackage{ulem}
\usepackage{soul}

\def\spose#1{\hbox to 0pt{#1\hss}}
\def\lta{\mathrel{\spose{\lower 3pt\hbox{$\mathchar"218$}}
     \raise 2.0pt\hbox{$\mathchar"13C$}}}
\def\gta{\mathrel{\spose{\lower 3pt\hbox{$\mathchar"218$}}
     \raise 2.0pt\hbox{$\mathchar"13E$}}}
\def\calP{{\mathcal P}}
\def\calR{{\mathcal R}}

\newcommand{\Mpl}{M_\mathrm{Pl}}
\newcommand{\de}[2]{\kern - #1 em \mathrm{d} #2}

\newcommand{\dd}{\mathrm{d}}

\newcommand{\Mpc}{\mbox{Mpc}}
\newcommand{\LF}{\left(}
\newcommand{\RF}{\right)}

\DeclareMathOperator{\erfc}{erfc}

\title{\boldmath Revisiting 
	primordial black holes formation from preheating instabilities: the case of Starobinsky inflation}

\author[a,b]{Daniel del-Corral}
\author[c]{Paolo Gondolo}
\author[d]{K. Sravan Kumar}
\author[a,b]{Jo\~ao Marto}

\affiliation[a]{Departamento de F\'{\i}sica, Universidade da Beira Interior, Rua Marqu\^{e}s D'\'Avila e Bolama 6200-001 Covilh\~a, Portugal}
\affiliation[b]{Centro de Matem\'atica e Aplica\c{c}\~oes da Universidade da Beira Interior, Rua Marqu\^{e}s D'\'Avila e Bolama 6200-001 Covilh\~a, Portugal}
\affiliation[c]{Department of Physics and Astronomy, University of Utah, Salt Lake City, UT 84112, USA}
\affiliation[d]{Institute of Cosmology and Gravitation, University of Portsmouth,
Dennis Sciama Building, Burnaby Road,
Portsmouth, PO1 3FX, United Kingdom}

\emailAdd{corral.martinez@ubi.pt}
\emailAdd{paolo.gondolo@utah.edu}
\emailAdd{sravan.kumar@port.ac.uk}
\emailAdd{jmarto@ubi.pt}

\abstract{ In recent years, the formation of primordial black holes (PBH) in the early universe inflationary cosmology has garnered significant attention. One plausible scenario for primordial black hole (PBH) formation arises during the preheating stage following inflation. Notably, this scenario does not necessitate any ad-hoc fine-tuning of the scalar field potential. This paper focuses on the growth of primordial density perturbation and the consequent possibility of PBH formation in the preheating stage of the Starobinsky model for inflation. The typical mechanism for PBH formation during preheating is based on the collapse of primordial fluctuations that become super-horizon during inflation (type I)  and re-enter the particle horizon in the different phases of cosmic expansion. In this work, we show that there exists a certain range of modes that remain in the sub-horizon (not exited) during inflation  (type II modes) but evolve identically to type I modes if they fall into the instability band, leading to large density perturbation above the threshold and can potentially also contribute to the PBH formation.  We detail the conditions determining the possible collapse of type I and type II modes whose wavelengths are larger than the Jeans length we derive from the effective sound speed of scalar field fluctuations.
Since the preheating stage is an 'inflaton' (approximately) matter-dominated phase, we follow the framework of the critical collapse of fluctuations and compute the mass fraction using the well-known Press-Schechter and the Khlopov-Polnarev formalisms, and compare the two. Finally, we comment on the implications of our study for the investigations concerned with primordial accretion and consequent PBH contribution to the dark matter. }

\makeatother
\makeatletter
\gdef\@fpheader{}
\makeatother

\begin{document} 
\maketitle
\flushbottom

%%%%%%%%%%%%%%%%%%%%%%%%%%%%%%%%
%% INTRODUCTION
%%%%%%%%%%%%%%%%%%%%%%%%%%%%%%%%

\section{Introduction}
\label{sec:intro}

Primordial black holes (PBH) were first proposed in 1967 by Zeldovich and Novikov \cite{Zeldovich:1967lct} and, independently, by Hawking and Carr in a series of articles \cite{Hawking:1971ei,Carr:1974nx,Carr:1975qj}. Soon after, the possibility that PBH could account for at least a part of the Dark Matter (DM) became an evident curiosity and possibility \cite{Chapline:1975ojl,Meszaros:1975ef}. Since PBH are supposed to form before nucleosynthesis, they can be considered non-baryonic DM candidates and therefore do not interfere with constraints on the baryonic abundance. Also, because of that, they can be formed with any mass. Those with masses below $\sim10^{15}g$ have probably evaporated via Hawking radiation  \cite{Hawking:1974rv} by now if we do not take into account the mere possibility of early clustering and accretion of small mass PBH into the heavy ones. 
\cite{DeLuca:2020jug,DeLuca:2021pls,Kritos:2020wcl,DeLuca:2022bjs}. 
PBH with masses bigger than $\sim10^{15}g$ are typically proposed as DM candidates, as generators of structure in the universe \cite{Meszaros:1975ef,Afshordi:2003zb} or even as seeds for the formation of supermassive black holes in the center of galactic nuclei \cite{Carr:1984id,Bean:2002kx}. See \cite{Villanueva-Domingo:2021spv,Carr:2020gox} for recent reviews on PBH. The PBH of small masses of range $0.1g-10^9g$, which can form from various mechanisms in the early Universe, might {already evaporated by now} (neglecting any primordial accretions). Specific processes of their evaporation might also be partly responsible for the abundant particle production in the early Universe, including the generation of matter-antimatter asymmetry observed in the Universe \cite{Garcia-Bellido:2019tvz,Carr:2019hud,Gondolo:2020uqv,Boudon:2020qpo}. Any early evaporation of PBH can leave their imprints via gravitational waves and it is also a possibility, motivated by several quantum gravity-based proposals, that there could still exist stable PBH remnants contributing to the present dark matter density \cite{Papanikolaou:2020qtd,Papanikolaou:2023crz,Domenech:2021wkk,Domenech:2023mqk}. The broader view is that PBH open new doors of investigations to probe physics at high energy or fundamental scales via numerous astrophysical and gravitational wave observations. 

Since the inflationary phase constitutes the most important phase of the primordial Universe \cite{Starobinsky:1980te,Starobinsky:1981vz,Guth:1980zm}, it is vital to understand if the PBH formation gets significant contribution within the scope of inflationary cosmology. 
Since the latest observations from cosmic microwave background (CMB) from Planck data \cite{Planck:2018jri} strongly favor the single-field inflationary scenario, it is adequate to restrict ourselves to the detailed study of primordial Universe with a single-field setup. Based on the available Planck data, Starobinsky and Higgs inflationary scenarios have become the favorite models as they fit so far, with the spectral index and tensor-to-scalar ratio constraints. The success of Statobinsky inflation, in particular, has gained a lot of attention because it is the first model of inflation in the modified gravity context, which has emerged from the foundations of quantum gravity    \cite{Starobinsky:1980te,Koshelev:2020xby}. After the release of Planck data,  Starobinsky-like models have become a basis for building UV-completions around them because exponentially flat potentials happen to explain more naturally the observation of near scale invariance of the CMB power spectra \cite{Kehagias:2013mya,Koshelev:2020xby}.
To have PBH formation, in the framework of single field inflation, the most explored scenario considers abrupt power spectrum enhancements in the last 30-40 e-folds of the regular 50-60 e-folds of initial expansion, which only happen if the inflationary potential contains several contrived slopes \cite{Garcia-Bellido:2017mdw,Carr:2020gox}. However, most of these studies are {primarily based on} phenomenological considerations, { since the theoretical motivation for specific shapes of the inflaton potential, which can accommodate PBH formation,} is still elusive \cite{Cole:2023wyx}. Another context in which PBH have been extensively investigated is during the stage of reheating and later epochs of radiation and matter-dominated phases \cite{Escriva:2019phb,Carr:2020gox}. However, right after the end of inflation, there exists a brief period of preheating, dominated by inflaton matter, a common feature in almost every single field model of inflation. 
This phase has been recently explored and projected to give small-scale PBH formation \cite{Jedamzik:2010dq,Jedamzik:2010hq,Martin:2019nuw,Martin:2020fgl}. The studies, as mentioned earlier, so far, only consider the possible collapse of (type I) modes that are superhorizon during inflation. Later, they can give rise to resonance instabilities during the preheating stage. The fate of the modes that experience preheating instabilities, entering from the sub-horizon evolution during inflation, is argued to be highly quantum mechanical. Thus, their contribution to PBH formation is an open question. 
%\tcb{However, in \cite{Lyth:2005ze}, the authors studied the production of PBHs for the subhorizon modes during a radiation-dominated universe upholding the practical applicability and providing the conceptual discussions in relation to the issues of quantum to classical transition of the fluctuations.  In that sense, if one considers the formation of a PBH, the mass function can be treated as a classical quantity. Our work instead focuses on the role of sub-horizon modes during inflation, which become unstable at the preheating stage.}
We call these modes type II modes, and our primary aim here is to study all the fluctuations that can experience preheating instabilities, which can potentially collapse and form PBH. The relevance of quantum-to-classical effects in generating preheating instabilities and collapse dynamics is important to mention. In the preheating stage, we may have a situation where classical modes can coexist with quantum ones. Whether quantum modes can trigger instabilities and contribute to classical collapse is a broader and non-trivial question.  Without any indication that this can be the case, one valid approach would be to ignore them. However, we choose not to do so to estimate the impact of type II modes for several quantities related to PBH formation.  
We identify a subclass of type II modes that, for all purposes, behave exactly as type I ones by growing and seeding instabilities during the preheating phase. To be more precise, our focus is on all the modes that can become unstable and lead to the universal growth of density perturbations.
Therefore, one motivating idea of this investigation is to revisit the preheating stage carefully and explore these type I and type II modes that evolve identically during preheating despite their evolution history during inflation. 
We choose the framework of Starobinsky inflation to study the preheating instabilities.
Overall, the study of preheating in association with PBH is significant because this phase precedes reheating, radiation, and regular matter-dominated eras where PBH formations have been widely investigated so far \cite{Ozsoy:2023ryl}. 

For an ideal matter-dominated universe, the effect of the pressure in stopping the collapse is not as important as in the radiation-dominated case \cite{Musco:2005}. In the context of inflationary preheating, understanding the effect of small non-zero pressure is crucial \cite{Sasaki:2018dmp,Hertzberg:2014iza,Cembranos:2015oya}. This stems from the fact that non-negligible pressure results in the non-zero effective sound speed for preheating density perturbations, which causes Jeans length to play a role in collapse dynamics.  
Another factor that comes into play in this case is the presence of non-spherical effects that would stop the collapse.  Khlopov and Polnarev \cite{Khlopov:1980mg,Polnarev:1985btg,Khlopov:1982ef,Khlopov:2008qy} pioneered the study of PBH production in a matter-dominated universe in the 80's in the context of grand unification, which was later refined in \cite{Harada:2016mhb,Harada:2017fjm} where essentially it was found that for a perturbation to collapse, it needs to be almost spherically symmetric, which indeed restricts the probability of formation. In the literature, the PBH formation has often been addressed in the context of perfect fluids with constant equations of state. The advantage of fluid approximations with the constant equation of state $w$ is that one can implement a critical collapse framework to study the over-densities above a threshold value $\delta_c(w)$ \cite{Carr:2020gox}. For example,
in \cite{Harada:2013epa} an analytical expression for the threshold as a function of $w$ is found and it is supported by the numerical results in \cite{Musco:2012au} for $w\ll1$. In the context of inflationary preheating \cite{Suyama:2004mz,Suyama:2006sr} studied the production of PBH during an exact matter-dominated phase of preheating after inflation, showing that a Tachyonic preheating could enhance the production of PBH. However, the preheating phase is approximately matter-dominated with an average equation of state $\langle w\rangle\ll 1$ due to oscillating inflaton. 

In our investigation, we closely follow  \cite{Martin:2020fgl}, where PBH during preheating instabilities has been studied in the context of chaotic inflation. It is found that the formation of PBH gets halted as preheating ends. That is when the inflaton field starts to decay into other particles, producing more radiation. Moreover, since this collapse is not instantaneous, when this collapse time is greater than the time available until the end of preheating, one can obtain a lower bound for the density contrast. In this work, we use this criterion and compute the scale-dependent threshold $\delta_c(k)$ to determine PBH masses from the critical collapse framework. Furthermore, we work with scalar field dynamics during preheating without perfect fluid approximations. Afterward, we evaluate the mass fraction for PBH during the preheating stage of Starobinsky inflation using the Press-Schechter and Khlopov-Polnarev methods.

The contents in this paper are organized as follows: In Sec.~\ref{sec:inflation-preheating}, we review the dynamics of the inflaton field during inflation and the consequent generation of curvature and density perturbations. Sec.~\ref{sec:numerics} describes the numerical procedure used to compute the background and perturbations equations, exploring some range of comoving wavenumbers $k$ meaningful to the density perturbations amplification. The process of PBH formation is explained in Sec.~\ref{sec:results} where, for the Starobinsky potential, we present numerical estimations of the PBH mass fraction and associated mass. Conclusions and outlook are presented in Sec.~\ref{sec:conclusions}. {Appendix~\ref{Appw} shows the computation of the averaged equation of state for an oscillating scalar field during preheating, respectively.} Appendices~\ref{appA} and \ref{appB} give details on how the initial conditions for the field and the perturbations are computed. 

Throughout the paper, we followed the metric signature $(-+++)$ and reduced Planck mass $\Mpl=\frac{1}{\sqrt{8\pi G}}$ with units of $\hbar=c=1$. 

%%%%%%%%%%%%%%%%%%%%%%%%%%%%%%%%
%% INFLATION PREHEATING AND DYNAMICS
%%%%%%%%%%%%%%%%%%%%%%%%%%%%%%%%

\section{Inflation and preheating dynamics}\label{sec:inflation-preheating}

\begin{figure}[h!]
	\centering    \includegraphics[scale=0.6]{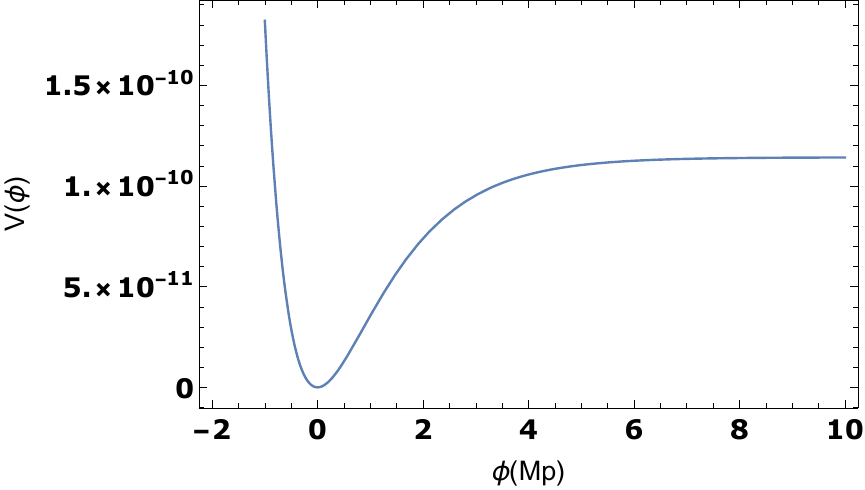}
	\caption{Plot of the potential \eqref{Staro}, with $V_0=1.14\times 10^{-10}$. This is normalized to match the amplitude of the power spectrum at the pivot scale, $P_\zeta(k_{pivot})=2.2\times10^{-9}$ \cite{Planck:2018jri}, where $k_{pivot}$ is given by $k_{pivot}=0.05\Mpc^{-1}=1.33\times 10^{-58}\Mpl$.}
	\label{fig:staro-pot}
\end{figure}

\begin{figure}
     \centering
     \begin{subfigure}[b]{0.49\textwidth}
         \centering         \includegraphics[width=\textwidth]{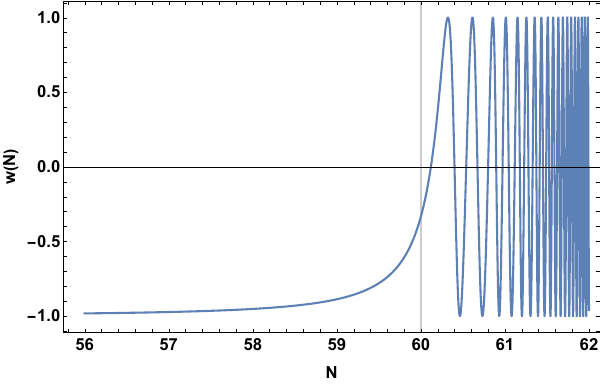}
         \caption{}
         \label{fig:EOS}
     \end{subfigure}
     \hfill
     \begin{subfigure}[b]{0.49\textwidth}
         \centering         \includegraphics[width=\textwidth]{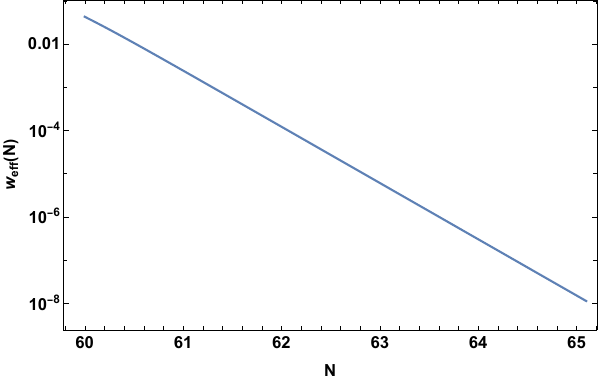}
         \caption{}
         \label{fig:EOS2}
     \end{subfigure}
        \caption{{a) The equation of state obtained numerically from \eqref{eq:EOS-full} b) The effective equation of state \eqref{eq:EOS-averaged} as a function of e-folds.}}
        \label{fig:equation-of-state}
\end{figure}

Let us begin with the background dynamics. In flat Friedmann-Lema\^itre-Robertson-Walker (FLRW) spacetime, the inflationary dynamics for the inflaton are given by the field equations
\begin{equation}
\label{KG}
\ddot{\phi}+3H\dot{\phi}+\frac{\dd V(\phi)}{\dd\phi} =0\,, 
\end{equation}
\begin{equation}
\label{FR}
    H^2 = \dfrac{V(\phi) + \dfrac{\dot{\phi}^2}{2}}{3\Mpl^2}\, \quad \dot{H} = -\frac{\dot{\phi}^2}{2}\,,
\end{equation}
Here, $\Mpl$ is the reduced Planck mass and $H=\frac{\dot{a}}{a}$ the Hubble parameter, with $a$ the scale factor of the universe and a dot meaning derivation with respect to cosmic time. The inflaton potential, according to Starobinsky's $R+R^2$ theory, is
\begin{equation}\label{Staro}
	V(\phi)=V_0\left(1-e^{-\sqrt{\frac23}\frac{\phi}{\Mpl}}\right)^2,
\end{equation}
where $V_0 = \frac{3}{4}M^2\Mpl^2$ is related to the 'scalaron' mass $M\sim 1.3\times 10^{-5}\Mpl$, in the Jordan frame, and $\phi$ is also expressed in units of Planck mass, $\Mpl$. The potential \eqref{Staro} is depicted in Fig.~\ref{fig:staro-pot}.
The scalar (density) perturbations evolution is controlled by the so-called Mukhanov-Sasaki variable $v_\textbf{k}$, whose equations motion is given by~\cite{Mukhanov:1990me,Baumann:2009ds}
\begin{equation}
\label{MSconformal}
v_{{\bm k}}''+\left[k^2-\frac{z''}
{z}\right]v_{{\bm k}}=0 \, .
\end{equation}
In this expression, a prime denotes derivative with respect to conformal time $\eta$, defined as $\dd t = a \dd \eta$, $z\equiv\sqrt{2\epsilon}\,a\Mpl$, where $\epsilon\equiv-\frac{\dot{H}}{H^2}$ is the first slow-roll parameter and the suffix $k$ denotes Fourier component. For practical reasons, working in cosmic time $t$ will be more efficient. Therefore, using the relation between cosmic and conformal time, Eq.\eqref{MSconformal} is now given by:
\begin{equation}\label{MScosmic}
    \Ddot{v}_k+H\dot{v}_k+\left[\frac{k^2}{a^2}+\frac{\dd^2 V}{\dd\phi^2}-2H^2+\frac{2\dot{\phi}}{H\Mpl^2}\frac{\dd V}{\dd\phi}+\frac{7\dot{\phi}^2}{2\Mpl^2}-\frac{\dot{\phi}^4}{2H^2\Mpl^4}\right]v_k=0 ,
\end{equation}
which is known as the Mukhanov-Sasaki equation. The reason behind this change is that all the terms in this equation remain non-singular after the end of inflation, when $\phi$, $\dot{\phi}$ and $\epsilon$ periodically vanish as the field oscillates around the bottom of the potential. Once we have solved for $v_k$, the Fourier component of the comoving curvature perturbation $\calR_k$ is related to the Mukhanov variable by
\begin{equation}\label{eq:zeta-v-relation}
    \calR_k=\frac{1}{\Mpl^2}\frac{v_k}{a\sqrt{2\epsilon}}.
\end{equation}
This is a very useful quantity since it allows us to compute the dimensionless power spectrum of curvature perturbations as 
\begin{equation}
    \mathcal{P}_\calR(k)=\frac{k^3}{2\pi^2}|\calR_k|^2.
\end{equation}
Let us see now how to compute the fractional energy density perturbations or density contrast, $\delta_k=\delta\rho_k /\rho$, where $\delta\rho_k$ is the Fourier component of the density perturbation and $\rho=3H^2\Mpl^2$ is the background energy density. To do so, we will start with the perturbed Einstein equations in cosmic time \cite{Baumann:2009ds}:
\begin{subequations}\label{eq:Einstein}
    \begin{equation}\label{eq:first-einstein}        3H(\dot\Psi+H\Phi)+\frac{k^2}{a^2}\left[\Psi+H(a^2\dot E-aB)\right]=-\frac{\delta\rho}{2\Mpl^2},
    \end{equation}
    \begin{equation}\label{eq:momentum-constraint}
        \dot\Psi+H\Phi=-\frac{\delta q}{2\Mpl^2}
    \end{equation}
    \begin{equation}\label{eq:second-einstein}       \ddot{\Psi}+3H\dot\Psi+H\dot\Phi+(3H^2+2\dot H)\Phi=\frac{1}{2\Mpl^2}\left(\delta p-\frac23k^2\delta\Sigma\right),
    \end{equation}
    
    \begin{equation}\label{eq:third-einstein}
        (\partial_t + 3H)(\dot E-B/a)+\frac{\Psi-\Phi}{a^2}=\frac{\delta\Sigma}{\Mpl^2},
    \end{equation}    
\end{subequations}
where $\delta q$ is the momentum density, $\delta p$ is the pressure perturbation, $\delta \Sigma$ is the anisotropic stress and $\Psi$, $\Phi$, $E$ and $B$ are the scalar metric perturbations, defined as 
\begin{equation}
    ds^2=-(1+2\Phi)dt^2+2aB_{,i}d x^id t+a^2\left[(1-2\Psi)\delta_{ij}+2E_{,ij}d x^id x^j\right].
\end{equation}
The comoving curvature perturbation is related to those quantities as \cite{Baumann:2009ds}
\begin{equation}
    \calR=\Phi-\frac{H}{\rho+p}\delta q.
\end{equation}
Using the perturbed Einstein equations, concretely the momentum constraint, which corresponds to \eqref{eq:momentum-constraint}, we obtain the following relation between the metric potentials $\Phi$ and $\Psi$ and the comoving curvature perturbation
\begin{equation}\label{eq:R-general}
    \calR=\frac23\frac{H^{-1}\dot\Psi+\Phi}{1+w}+\Psi,
\end{equation}
where $w$ is the equation of state, defined as
\begin{equation}\label{eq:EOS-full}
    w=\frac{p}{\rho}=\frac{\frac{\dot{\phi^2}}2-V(\phi)}{\frac{\dot{\phi^2}}2+V(\phi)}.
\end{equation}
Fig.~\ref{fig:EOS} shows the equation of state for the Starobinsky model. During inflation, it reaches a constant value of $w=-1$. However, during preheating, it oscillates around zero, and one needs to define an effective equation of state. In Appendix.~\ref{Appw}, we show the computation procedure of the effective equation of state (see Eqn.~\eqref{eq:EOS-averaged}), which is plotted in Fig.~\ref{fig:EOS2}. This shows that preheating is nearly a matter-dominated stage due to the smallness of the effective equation of state.
From now on, we will work in the Newtonian gauge to take $E=B=0$. Moreover, since we are dealing with a scalar field, the anisotropic stress vanishes and, therefore, \eqref{eq:third-einstein} implies that $\Phi=\Psi$. Using this into \eqref{eq:R-general} we have 
\begin{equation}\label{eq:comoving-curvature}
    \calR_k=\frac23\frac{H^{-1}\dot\Phi_k+\Phi_k}{1+w}+\Phi_k.
\end{equation}
 If we now take \eqref{eq:first-einstein} and divide it by $3H^2$ we obtain
\begin{equation}\label{eq:density-pert}
    \delta_k=-\frac23\left(\frac{k^2}{a^2H^2}+3\right)\Phi_k-2\frac{\dot\Phi_k}{H},
\end{equation}
where, again, the Fourier component has been considered. In this expression we express the density perturbation $\delta_k$ in terms of the perturbation $\Phi_k$, which is obtained by solving the differential equation \eqref{eq:comoving-curvature}, where $\mathcal{R}_k$ is given by \eqref{eq:zeta-v-relation} and \eqref{MScosmic}. 

\begin{figure}
     \centering
     \begin{subfigure}[b]{0.49\textwidth}
         \centering         \includegraphics[width=\textwidth]{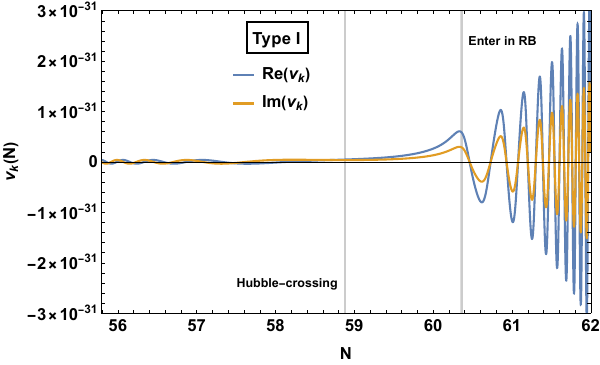}
         \caption{}
         \label{fig:compare-type-I}
     \end{subfigure}
     \hfill
     \begin{subfigure}[b]{0.49\textwidth}
         \centering         \includegraphics[width=\textwidth]{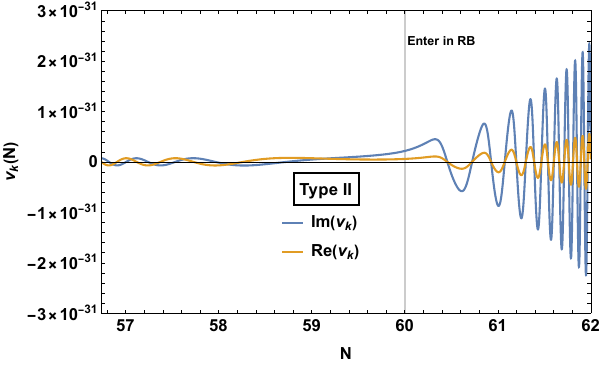}
         \caption{}
         \label{fig:compare-type-II}
     \end{subfigure}
     \begin{subfigure}[b]{0.49\textwidth}
         \centering         \includegraphics[width=\textwidth]{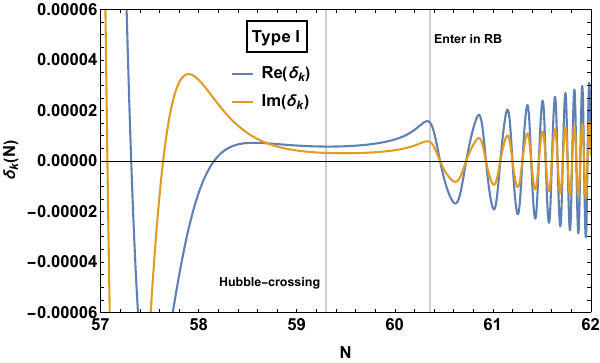}
         \caption{}
         \label{fig:compare-type-I-density}
     \end{subfigure}
     \hfill
     \begin{subfigure}[b]{0.49\textwidth}
         \centering         \includegraphics[width=\textwidth]{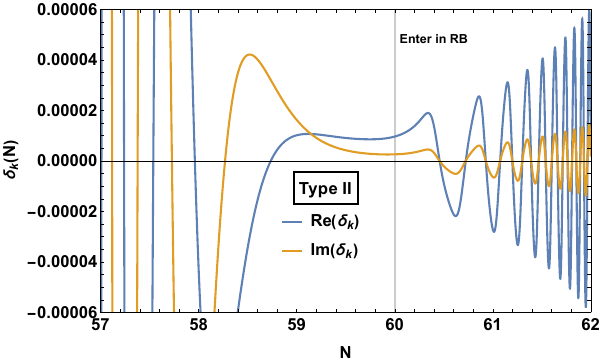}
         \caption{}
         \label{fig:compare-type-II-density}
     \end{subfigure}
        \caption{{a) Time evolution of the Mukhanov-Sasaki variable for a type I mode ($k\simeq0.7 k_{\text{end}}$). Vertical gray lines mark the points where the mode exits the horizon during inflation (left one) and where it reenters back (right one). b) The same but for a type II mode ($k=1.35 k_{\text{end}}$). Vertical gray line marks, in this case, the point where the mode enters the resonance band, and thus, it starts to amplify. c) Time evolution of the density contrast for the same type I mode. d) Time evolution of the density contrast for the same type II mode.}}
        \label{fig:compare-I-and-II}
\end{figure}

Equation \eqref{eq:comoving-curvature} can be solved numerically, which requires some suitable initial conditions for $\Phi_k$. These are obtained by considering the behavior of the curvature perturbation $\mathcal{R}_k$ in two different regimes of $k$. The first one (type I modes) is defined by the modes that exited the horizon during inflation (and, therefore, enter the particle horizon during preheating). The second (type II modes) involves wavenumbers that have never exited the horizon but become relevant as type I when they enter the particle horizon during preheating. For the former ones, we can define the last scale to enter the Hubble radius during preheating, $k_{\text{min}}$, which, using the Hubble radius crossing condition, can be computed as 
\begin{equation}
k_{\text{min}}=a(t_{\text{rh}})H(t_{\text{rh}}),
	\label{eq:kmin}  
\end{equation}
where $t_{\text{rh}}$ is the time at which preheating ends. Since the preheating duration depends on the inflaton coupling to matter fields and its decay process, we will consider various arbitrary periods of preheating. Numerically, this means to choose a sufficiently small $k_{\text{min}}$ such that it never crosses the horizon during preheating, and thus we can study all modes that enter the horizon during this phase.  Also, the last scale to exit the horizon during inflation, $k_{\text{end}}$, is defined as 
\begin{equation}\label{eq:kend} k_{\text{end}}=a(t_{\text{end}})H(t_{\text{end}}),
\end{equation}
where $t_{\text{end}}$ is the time at which inflation ends {which we consider to be at $N=60$ and same followed throughout the rest of the paper}. These type I modes belong to the interval $k\in[k_{\text{min}},k_{\text{end}}]$. {Type II modes are those wavenumbers bigger than type I and never exit the horizon during inflation. These modes are not considered in the literature because of their sub-horizon (quantum-mechanical) evolution during inflation. Since they never leave the horizon during inflation, it is usually considered that those modes are still in the quantum regime.
Our investigation reveals that despite their history, a subclass of type II modes during preheating gets amplified like type I.   The collapse of type I modes is considered as they enter the particle horizon and fall into the instability band during preheating.  Meanwhile, the subclass of type II modes never exits the horizon during inflation but can still fall into the instability band and get amplified. Therefore,
both type I and II modes are on equal footing (in the instability band). Thus, we find both types I and II equally contribute to the PBH formation, as explained in the later part of the section. 
For illustration, in Fig.~\ref{fig:compare-I-and-II}, we have plotted the evolution of the Mukhanov-Sasaki variable $v_k$ and the density contrast $\delta_k$ for a type I and a type II mode, both real and imaginary parts. We assumed the standard Bunch-Davies initial conditions described in Appendix \ref{appB}. 
The question of the quantum aspects of these modes that could potentially collapse remains, we nonetheless explore all the modes classically in this work leaving the quantum treatment for future investigations.} 

\begin{figure}
    \centering
    \includegraphics[scale=1]{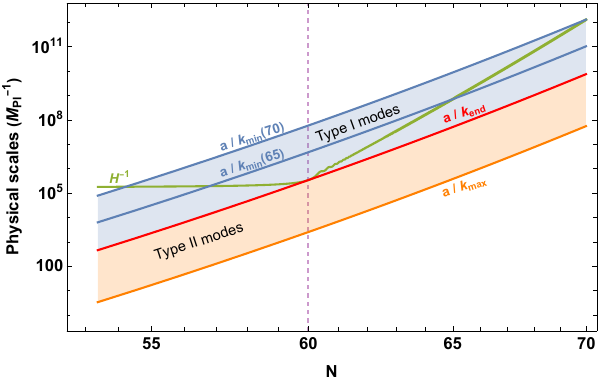}
    \caption{Physical scales of interest (see \eqref{eq:kphys}) for the Starobinsky model as well as the physical Hubble radius $H^{-1}$ as a function of the number of e-folds from the beginning of inflation. The small oscillations in $H^{-1}$ after $N=60$ are part of the solution, not an artifact. The number inside the parenthesis in $k_{\text{min}}$ indicates the number of e-folds at which they re-enter the horizon. The Blue shaded area corresponds to the type I modes, whereas the orange shaded area corresponds to the type II modes.}.
    \label{fig:scales}
\end{figure}

\begin{figure}
    \centering
    \includegraphics[scale=0.9]{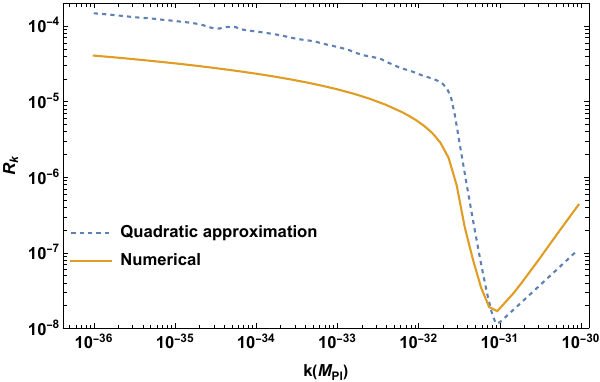}
    \caption{{Comparison between the quadratic approximation $V\sim\frac{M^2}2\phi^2$ using Floquet theory and the numerical solution of the full Starobinsky potential \eqref{Staro} when computing the curvature perturbations. Evaluation is made at $N=65$, which is 5 e-folds after the end of inflation at $N=60$. Although the modes that get excited are the same, the Floquet exponents are not.}}
    \label{fig:compare-floquet}
\end{figure}

\begin{figure}
     \centering
     \begin{subfigure}[b]{0.49\textwidth}
         \centering         \includegraphics[width=\textwidth]{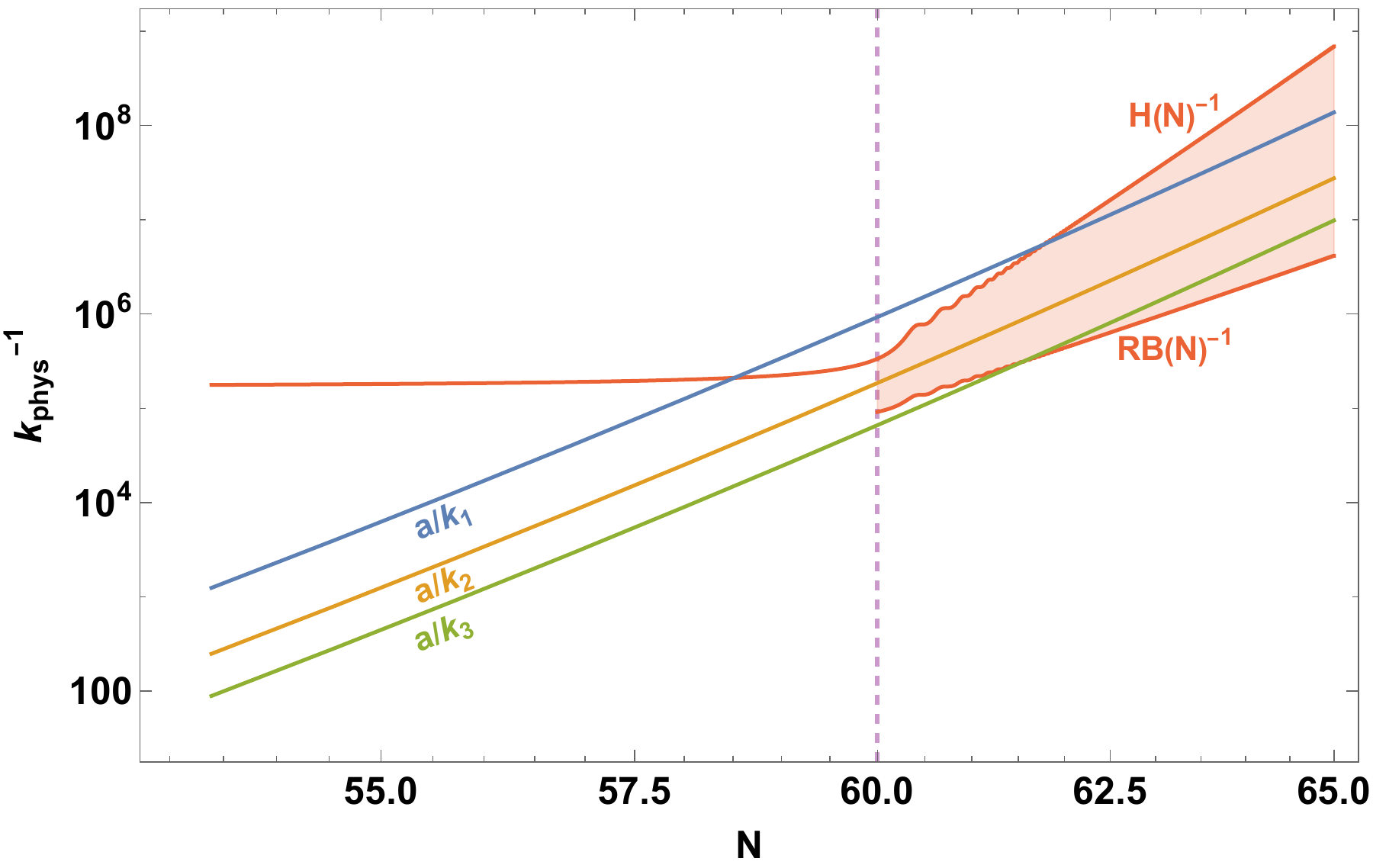}
         \caption{}
         \label{fig:parametric-instability}
     \end{subfigure}
     \hfill
     \begin{subfigure}[b]{0.49\textwidth}
         \centering         \includegraphics[width=\textwidth]{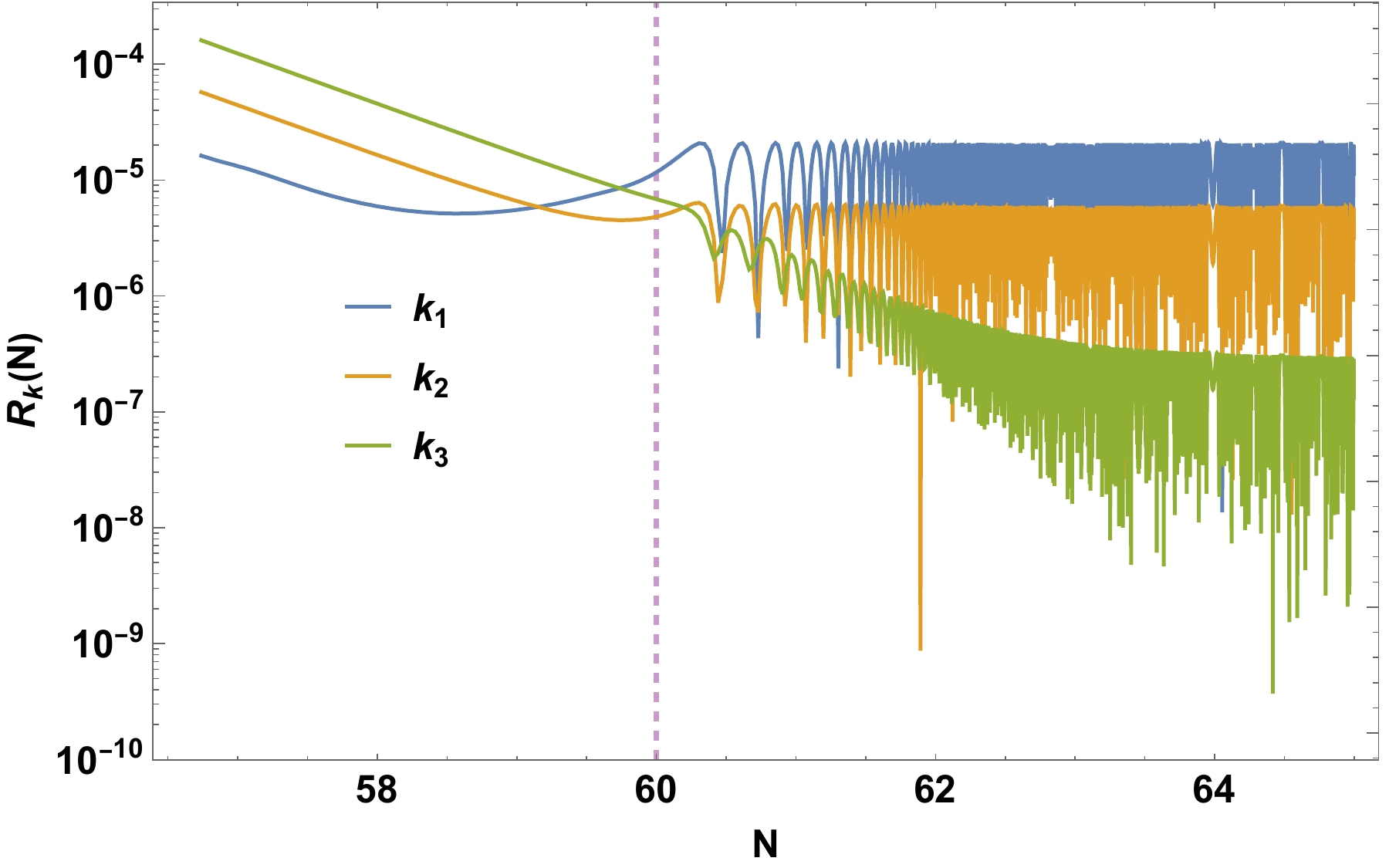}
         \caption{}
        \label{fig:exmaple-scales}
     \end{subfigure}
        \caption{\textbf{a)} Resonance band (red shaded area) \eqref{eq:RB} as well as some examples of physical scales during the end of inflation and the beginning of preheating. \textbf{b)} Curvature perturbation for the physical scales selected. The vertical purple dashed line marks the end of inflation. The $\epsilon$ parameter is set to $\epsilon=3/2$ for visual purposes due to its high oscillating behavior.}
        \label{fig:resonance-band}
\end{figure}

Now, for the type II modes, we fix an upper limit $k_{\text{max}}$ (based on our numerical evaluation of density perturbations for the Starobinsky model), given by
\begin{equation}\label{eq:kmax}
    k_{\text{max}}=10^{-30}\Mpl,
\end{equation}
where the density perturbations typically reach the non-perturbative regime, at least for the Starobinsky model, see Fig.~\ref{fig:alldensities-staro}. These type II modes belong to the interval $k\in[k_{\text{end}},k_{\text{max}}]$. In Fig.~\ref{fig:scales}, all these scales are represented for different periods of the preheating stage, as well as the Hubble radius $H^{-1}$. Bear in mind that $H^{-1}$ is a physical scale (in contrast to the conformal scale $(aH)^{-1}$) and, therefore, when plotted together with the scales k, those scales must be physical too. We define physical scales as
\begin{equation}\label{eq:kphys}
    k_{\text{phys}}=\frac{k_{\text{com}}}{a},
\end{equation}
where $k_{\text{com}}$ is the comoving wavenumber, whose values for the scales of interest are given by Eqns.~\eqref{eq:kmin}-\eqref{eq:kend}.
Now, before solving \eqref{eq:comoving-curvature} it is worth mentioning the effect of parametric resonance. This will help us to have a better picture of how the curvature perturbations behave during preheating. Since, during this phase, the inflaton field oscillates around the bottom of the potential \eqref{Staro}, one can approximate it by a quadratic one, that is, $V\sim\frac{M^2}{2}\phi^2$. Doing so makes it possible to rewrite the Mukhanov-Sasaki equation \eqref{MScosmic} as a Mathieu-like equation of the type.
\begin{equation}
    \frac{\dd^2\tilde{v}_k}{\dd z^2}+\left[A_k-2q(z)\cos{(2z)}\right]\tilde{v}_k=0,
\end{equation}
where $\tilde{v}_k=a^{1/2}v_k$ is the re-scaled Mukhanov-Sasaki variable,  $z=Mt+\pi/4$ and the parameters $A_k$ and $q$ depend on the background solution \cite{Martin:2019nuw,Martin:2020fgl}. Now, since $q\ll1$ we are in the narrow resonance regime, and the first resonance band (RB) is given by the condition $1-q<A_k<1+q$, which corresponds to
\begin{equation}
    0<\frac{k}{a}<\sqrt{3HM}\frac{M\phi_{\text{end}}}{\sqrt{6}H_{\text{end}}\Mpl}.
\end{equation}
If we now consider physical scales (in length units) and the fact that the mode has to be inside the horizon so that causality applies, we have that for a mode to be in the RB means the following
\begin{equation}\label{eq:RB}
    \frac1{H}>\frac{a}{k}>\frac{1}{\sqrt{3HM}}\frac{\sqrt{6}H_{\text{end}}\Mpl}{M\phi_{\text{end}}}\simeq\frac{1}{\sqrt{3HM}}.
\end{equation}
{The important thing here is the fact that for the modes satisfying \eqref{eq:RB}, the re-scaled Mukhanov-Sasaki variable grows as {$\tilde{v}_k\simeq e^{\int\mu(z)\dd z}\simeq a^{3/2}$, where $\mu(t)\sim\frac{q(z)}2$ is the so-called Floquet exponent. All this} is equivalent to $v_k\simeq a$. This implies a constant curvature perturbation by the definition of \eqref{eq:comoving-curvature}. That is: \textit{modes satisfying \eqref{eq:RB} (inside the resonance band) have a constant curvature perturbation}. Fig.~\ref{fig:resonance-band} confirms numerically this fact. We depict the resonance band in Fig.~\ref{fig:parametric-instability} with some examples of physical scales. In Fig.~\ref{fig:exmaple-scales}, we depict the curvature perturbation associated with those physical scales as a function of the number of e-folds. For the type I mode labeled as $k_1$, the curvature perturbation decays during inflation until it exits the horizon and gets fixed to a constant value even after entering the horizon (or, equivalently, entering the RB). We also observe the curvature perturbation for type II modes $k_2$ and $k_3$, which enter the resonance band at different times. The mode $k_2$ decays until the end of inflation, and as it enters the resonance band at this point, it evolves towards constant value. On the other hand, the mode $k_3$ decays until approximately $N=62$, then it enters the RB and eventually approaches a constant value.} {It is important to mention that approximating the potential \eqref{Staro} by a parabola reflects the true behavior of the Starobinsky model for prolonged preheating stages. However, during the first moments of preheating, the Starobinsky potential differs from the quadratic one (see Fig.~\ref{fig:potentials} for high values of $\phi$). Besides, we can notice that the expression for the resonance band \eqref{eq:RB} accurately reflects the modes that become unstable. To better understand the importance of not considering the Starobinsky potential as a parabola, we have followed the steps of \cite{Martin:2019nuw,Martin:2020fgl}. We have computed the Floquet exponents for the quadratic potential $V\sim\frac{M^2}2\phi^2$, with $M$ being the scalaron mass of the Starobinsky model\footnote{{Notedly the mass of the inflaton in the quadratic inflation is 10 times smaller than the scalaron of Starobinsky model \cite{Harigaya:2015pea}. Due to this, we see few differences in our results of resultant PBH masses compared to those obtained in  \cite{Martin:2019nuw,Martin:2020fgl} in the context of chaotic inflation.} }. Using these exponents, one can compute the curvature perturbation $\mathcal{R}_k$. The results are displayed in Fig.~\ref{fig:compare-floquet}, where we show the comparison between the computation of the curvature perturbation using the parabola approximation for the potential together with Floquet theory vs. the numerical computation using the full expression \eqref{Staro}. Evaluation is made at $N=65$, and one can notice that the approximation does not seem to reproduce the numerical results accurately. This is mainly because, during the early stages of preheating, the Floquet exponents for the quadratic approximation are higher than those computed using the full potential. We, therefore, remark on the importance of considering the higher order terms than the quadratic in this computation\footnote{{Worth to note here that the quadratic potential approximation becomes even more inaccurate for a more general class of potentials like $\alpha-$ attractors \cite{alphaatr,self-resonance} }.}. 
For a detailed computation of Floquet's exponents, including higher order terms than the quadratic one, see \cite{self-resonance}. 
For more details about the computation of perturbation, see the following sections.}
Now that we have specified the two ranges of k and understood the behavior of the curvature perturbation during preheating, we can solve \eqref{eq:comoving-curvature} analytically. For type I, since the modes exit the horizon during inflation, the curvature perturbation becomes and remains nearly constant even when the mode re-enters the particle horizon during preheating since it enters the RB, see Fig.~\ref{fig:resonance-band}.
We can thus fix the value of $\calR_k$ to $\calR_k^{\text{hc}}\sim\calR_k^{\text{end}}$, the magnitude it had when it crossed the Hubble scale (hc stands for Hubble crossing). Therefore, \eqref{eq:comoving-curvature} can be solved analytically to give
\begin{equation}\label{eq:bardeen-super}
    |\Phi_k|=\frac35|\calR_k^{\text{end}}|+C_1(k)\left(\frac{a_{\text{end}}}{a}\right)^{5/2},
\end{equation}
where $C_1(k)$ is a constant of integration. Ignoring the decaying mode, we see that the perturbation $\Phi_k$ also remains constant during preheating in this interval. To obtain \eqref{eq:bardeen-super}, we have considered that $H\sim\frac2{3t}$ and that {$w_{\text{eff}}\sim0$ during preheating, where the angular brackets mean averaging over one period of oscillation. See Appendix~\ref{Appw} and Fig.~\ref{fig:equation-of-state} for a discussion of how the averaged equation of state is computed}. Regarding the derivative of $\Phi_k$, by direct inspection of Eqn. \eqref{eq:comoving-curvature} one can deduce that $|H^{-1}\dot{\Phi}_k|\ll |\Phi_k|$, so that $|\Phi_k|=\frac35|R^{\text{end}}_k|$ applies.
%\begin{figure}
%     \centering
%     \begin{subfigure}[b]{0.49\textwidth}
 %        \centering         \includegraphics[width=\textwidth]{Images/omega.pdf}
   %      \caption{}
   %      \label{fig:omega}
   %  \end{subfigure}
   %  \hfill
   %  \begin{subfigure}[b]{0.49\textwidth}
   %      \centering         \includegraphics[width=\textwidth]{Images/omega-mean.pdf}
   %      \caption{}
   %     \label{fig:omega-average}
   %  \end{subfigure}
   %     \caption{\tcr{In this figure, we have the equation of state for the Starobinsky model. a) Numerical solution and b) time-averaged value of $w$. During inflation, we see that $w\sim-1$ and during preheating $w\sim0$ (on average). The averaging procedure is made as follows: for each value of $N$, we compute the average of the values of $w$ inside the interval $[N-0.1, N+0.1]$.}}
%        \label{fig:omega-num-average}
%\end{figure}
Using this into \eqref{eq:density-pert} (ignoring the decaying term), we have that the density perturbation during preheating and for the type I modes is given by
\begin{equation}\label{eq:density-pert-sup}
    |\delta_k^{I}|\sim\frac25\left(\frac{k^2}{a^2H^2}+3\right)|\calR_k^{\text{end}}|.
\end{equation}
If the mode is super-Hubble, the term $k^2/a^2H^2$ is negligible, and the density perturbation is constant. However, once the mode enters the particle horizon during preheating, this term grows as
\begin{equation}\label{eq:dominant-term}
    \frac{k^2}{a^2H^2}\sim\left(\frac{k}{k_{\text{end}}}\right)^2\left(\frac{a}{a_{\text{end}}}\right),
\end{equation}
which implies the growth of density perturbation as $\delta_k\sim a$ \cite{Martin:2020fgl}. This increase will depend on both the value of $k$ with respect to $k_{\text{end}}$ and $a$ with respect to $a_{\text{end}}$. For example, if the mode enters the particle horizon at the beginning of preheating, then $k\sim k_{\text{end}}$, and therefore the increase will be maximum. However, if the mode enters the particle horizon by the end of preheating $k\ll k_{\text{end}}$, then the increase will be minimal. Therefore, what one can say here is that among the type I modes, those that enter the particle horizon $k \approx k_{end}$ are likely to create overdense regions that can potentially collapse and form PBH. We can call these type I modes those that experience instability and are the modes partly studied earlier \cite{Martin:2019nuw}. Not all overdensities in the preheating stage can collapse contrary to the assumptions made earlier  \cite{Martin:2019nuw}. We shall discuss later in detail that the threshold energy density and Jeans instability criterion plays a crucial role in pinning down the modes likely to collapse. These considerations suggest we revisit the previous estimates of PBH formation during the preheating instabilities \cite{Martin:2019nuw}.

Let us now study type II modes. In this case, the Mukhanov-Sasaki variable is still approximately given by the vacuum solution during inflation (see Appendix \ref{appB} or equation \eqref{BDconformal}). Suppose the modes enter into the resonance band during preheating. In that case, the curvature perturbation remains constant (see \cite{Martin:2019nuw,Martin:2020fgl} for details), and we retrieve solutions \eqref{eq:bardeen-super} and \eqref{eq:density-pert-sup} again. This would be the case of mode $k_2$ in Fig.~\ref{fig:resonance-band}, the mode $k_3$ after $N\sim62$ or, in general, any type II mode that falls into the RB.
That is, the density perturbation grows as
\begin{equation}\label{eq:density-pert-sub}
    |\delta_k^{II}|\sim\frac25\left(\frac{k^2}{a^2H^2}+3\right)|\calR_k^{\text{end}}|.
\end{equation}
The modes that are of interest to us are {the type I and type II modes that fall into the RB
\begin{equation}
    \delta_k^{II}\sim\delta_k^{I} \sim a.
\end{equation}
Type II modes outside the RB\footnote{The curvature perturbation \eqref{eq:zeta-v-relation} decays as $\sim a^{-1}$ if the mode is outside the RB (see Fig.~\ref{fig:resonance-band} for mode $k_3$ before $N\sim62$). This means it can be expressed as
\begin{equation}  \label{eq:curvature-sub-horizon}  |\calR_k|\sim\left(\frac{a_{\text{end}}}{a}\right)|\calR_k^{\text{end}}|.
\end{equation}
Using \eqref{eq:curvature-sub-horizon} into \eqref{eq:comoving-curvature} and solving for $\Phi_k$ we obtain
\begin{equation} \label{eq:bardeen-sub}   |\Phi_k|\sim\left(\frac{a_{\text{end}}}{a}\right)|\calR_k^{\text{end}}|+C_2(k)\left(\frac{3H}{2}\right)^{5/2},
\end{equation}
where again $C_2(k)$ is a constant of integration. Now, by looking at Eqn. \eqref{eq:comoving-curvature} we see that $H^{-1}\dot{\Phi}_k+\Phi_k\simeq 0$, which implies $\calR_k\simeq\Phi_k$ (Eqn. \eqref{eq:bardeen-sub}). Using this into \eqref{eq:density-pert} and ignoring the term decaying as $\sim H^{5/2}$, the density perturbation in this interval and for the modes whose curvature perturbation decays as $a^{-1}$ (that is, modes outside the RB) is approximately given by
\begin{equation}\label{eq:density-pert-sub-II}
    |\delta_k^{II}|\sim\frac23\frac{k^2}{a^2H^2}\left(\frac{a_{\text{end}}}{a}\right)|\calR_k^{\text{end}}|
    \sim\frac23\left(\frac{k}{k_{\text{end}}}\right)^2|\calR_k^{\text{end}}|,
\end{equation}
In the last step, we have used \eqref{eq:dominant-term}. Equation \eqref{eq:density-pert-sub-II} tells us that the density perturbations corresponding to the type II modes outside the RB remain constant. It is fixed by the value of curvature perturbation at the end of inflation and the quantity $(k/k_{\text{end}})^2$. For $k\gg k_{\text{end}}$, the value of $\delta_k$ can be so much larger that we require to go beyond the linear regime. We then exclude these values of $\delta_k$, as stated before. Furthermore, in the later sections we show that the modes with very small wavelengths (high k) fall shorter than the Jeans length and eventually unimportant for collapse dynamics. } are unimportant as their possible collapse is halted by Jeans criterion.}

Now that we know the relation between $\calR_k$ and $\delta_k$ we can use \eqref{eq:density-pert-sup} and \eqref{eq:density-pert-sub} to obtain the density perturbations. This is done by numerically solving for $\calR_k$, shown in the next section.

%%%%%%%%%%%%%%%%%%%%%%%%%%%%%%%

\section{Numerical approach}\label{sec:numerics}

%To compute the numerical evolution of the background and perturbations equations we have used the software \textit{Mathematica}. 
In this section, we present the numerical procedure for dealing with the background and, subsequently, with perturbations equations. Initial conditions and their justification can be found in Appendices \ref{appA} and \ref{appB}.
%%%%%%%%%%%%%%%%%%%%%%%%%%%%%%%

\subsection{Background}

\begin{figure}
    \centering
    \includegraphics[scale=0.6]{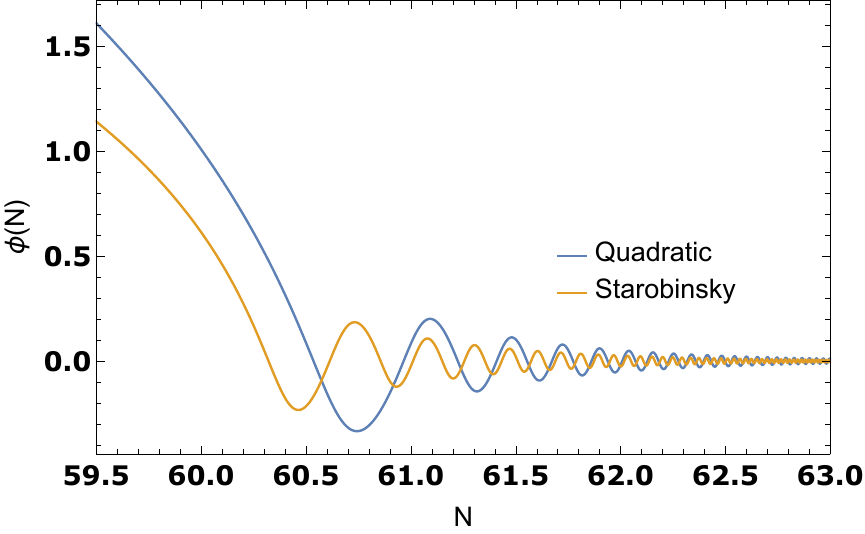}
    \caption{Comparison between quadratic and Starobinsky model during preheating.}
    \label{fig:quad-staro}
\end{figure}

First, the background equations have to be solved. Usually, one substitutes \eqref{FR} into \eqref{KG} so that there is just one second-order ordinary differential equation at the end. This kind of equation needs two initial conditions to be completely solved, one for $\phi$ and one for $\dot{\phi}$. We will derive the initial conditions from the slow-roll approximation. For the quadratic model, the expressions are given by
\begin{equation}\label{initialQuad}
    \phi\simeq 2\sqrt{N_{\text{inf}}}\Mpl\,,\qquad
    \frac{\dd\phi}{\dd N}\simeq-\frac{\Mpl}{\sqrt{N_{\text{inf}}}}\,, \qquad \epsilon\simeq\frac1{2N_{\text{inf}}}\,,
\end{equation}
where $N_{\text{inf}}$ stands for the total number of e-foldings of the scale factor during inflation. To translate derivatives with respect to $N$ to time derivatives, we can use the relation\footnote{This means that we need to multiply by $H$ when translating from $N$ time to $t$ time. The Hubble rate $H$ has to be evaluated also at $N=N_{\text{inf}}$ (or equivalently at $t=t_i$), which can be obtained from the normalization of the power spectrum at the pivot scale$$\mathcal{P}_\zeta(k_{\text{pivot}})\simeq2.2\times10^{-9}\simeq\frac{H^2}{8\pi^2\Mpl^2\epsilon(N_{\text{inf}})}$$}: $\dd N=\dd\ln{a}=H\dd t$. We will choose $N_{\text{inf}}=60$, as it is usually done. For the Starobinsky model with potential given by \eqref{Staro} instead, we have the following expressions
\begin{equation}\label{initialStaro}
    e^{\sqrt{\frac23}\frac{\phi}{\Mpl}}\simeq\frac{4N_{\text{inf}}}3\,, \qquad
    \frac{\dd\phi}{\dd N}\simeq-\sqrt{\frac{2}{3}}\frac{\Mpl}{N_{\text{inf}}} \,,\qquad
    \epsilon\simeq\frac3{4N_{\text{inf}}^2}.
\end{equation}
See Appendix \ref{appA} for a complete derivation of these expressions. We can now use  Eqns.\eqref{initialQuad} and \eqref{initialStaro} as initial conditions for the background equations \eqref{FR} and \eqref{KG}. Even though our computations are made in cosmic time, it is more intuitive to make the plots with respect to the number of e-folds. Therefore, we use the following relation 
\begin{equation}
    N(t)=\ln{\frac{a(t)}{a_0}},
\end{equation}
where $a_0$ is the scale factor at the beginning of inflation. This is evaluated by using the Hubble-cross condition at the pivot scale ($k_{\text{pivot}}=0.05\Mpc^{-1}=1.33\times 10^{-58}\Mpl$), that is:
\begin{equation}\label{eq:norm}
    a_0=\frac{k_{\text{pivot}}}{H(t_0)},
\end{equation}
Where $t_0$ is the time corresponding to 60 e-folds before the end of inflation. Notice that we should not confuse $t_0$ with the initial time $t_i$\footnote{The times $t_i$ and $t_0$ do not necessarily need to coincide. Imposing initial conditions from the slow-roll approximation gives an approximate (but somewhat accurate) estimation of the initial conditions for the field and seldom produces exactly 60 e-folds of inflation. This means that imposing initial conditions at $t_i$ can give more or less than 60 e-folds, but then we normalize the scale factor at $t_0$ so that from $t_0$ to the end of inflation, we have exactly 60 e-folds.}, considered in our computations. If this cosmic time is expressed in Planck mass units, it is enough to take $t_i=10^5\Mpl^{-1}$ to be accurate, but numerically there is no difference so that we can take $t_i=1\Mpl^{-1}$ as initial computational time. The scale factor $a(t)$ can be obtained by solving Friedmann equation \eqref{FR} with an arbitrary initial condition. After that, it is just a matter of defining a rescaling using \eqref{eq:norm} at $t_0$. Fig.~\ref{fig:quad-staro} shows for comparison both quadratic and Starobinsky models during the end of inflation and preheating, showing that in the latter case, the field decays faster than for the former one.

%%%%%%%%%%%%%%%%%%%%%%%%%%%%%%%%

\subsection{Perturbations}

As previously explained, we study perturbations by using the Mukhanov-Sasaki equation written in cosmic time \eqref{MScosmic}. Following \cite{Ballesteros:2017fsr}, it is better (for numerical purposes) to solve this equation separately for the real and imaginary parts of the Mukhanov variable. This means we must specify real and imaginary initial conditions when solving Eqn.\eqref{MScosmic}. For the scalar perturbations case, a common choice of initial conditions is given by the Bunch-Davis vacuum, which amounts to
\begin{equation}
\lim_{k/(aH)\rightarrow +\infty}v_{k}(\eta)=\frac{1}{\sqrt{2k}}
{\rm e}^{-ik\eta}\, ,
\label{BDconformal}
\end{equation}
when the mode is deep inside the Hubble radius ($k\gg aH$). Since we are working in cosmic time, these initial conditions must also be changed. The result is:
\begin{equation}\label{eq:ic-pert}
    \operatorname{Re}\left[v_k(t_i)\right]=\frac{1}{\sqrt{2k}}, \qquad
    \operatorname{Re}\left[\dot{v}_k(t_i)\right]=0, \qquad
    \operatorname{Im}\left[v_k(t_i)\right]=0, \qquad 
    \operatorname{Im}\left[\dot{v}_k(t_i)\right]=-\frac{1}{a}\sqrt{\frac{k}{2}},
\end{equation}
Where $t_i$ is the initial computational cosmic time. See Appendix \ref{appB} for a complete derivation of these formulae and initial time $t_i$. Using \eqref{eq:ic-pert} as initial conditions for \eqref{MScosmic} allows us to solve for the Mukhanov variable $v_k$ and then for the curvature perturbation $\calR_k$, defined in \eqref{eq:zeta-v-relation}. 
\begin{figure}
    \centering
    \includegraphics[scale=0.35]{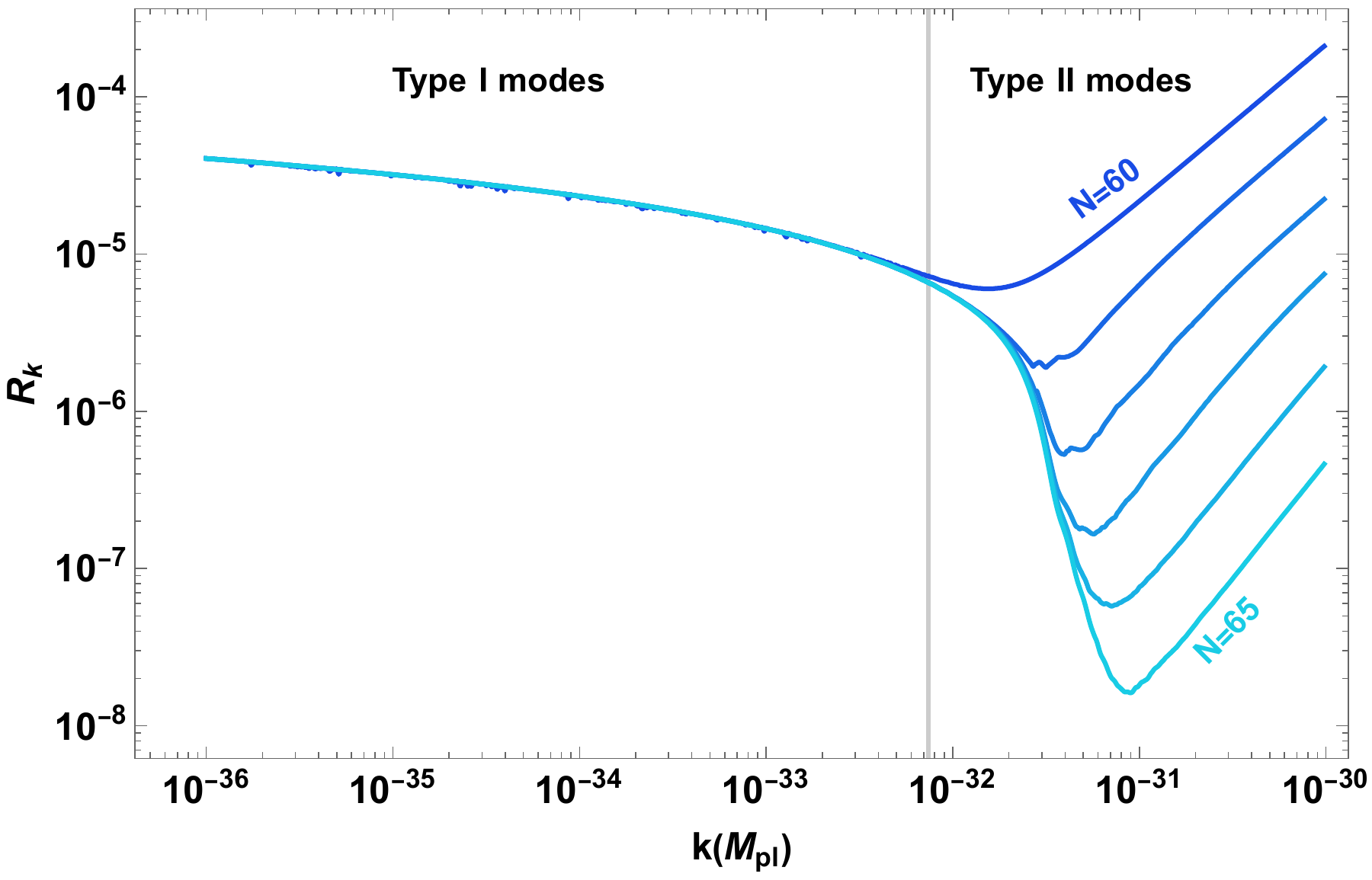}
    \caption{Averaged values of curvature perturbation as a function of the comoving wavenumber $k$ evaluated from $N=60$ (darker blue) to $N=65$ (lighter blue) and normalized using the factor $k^{3/2}$. The vertical grey line at $k_{\text{end}}$ separates the type I from the type II modes. This plot is entirely evaluated numerically.}
    \label{fig:allcurvatures-staro}
\end{figure}
Fig.~\ref{fig:allcurvatures-staro} shows the evolution of the comoving curvature perturbation for the two types of modes. In figures \ref{fig:allcurvatures-staro} and \ref{fig:alldensities-staro}, time grows from the curve marked as $N=60$ to the curve $N=65$. 
Also, all quantities represented in the figures are averaged and normalized with the factor $k^{3/2}$ \cite{Mukhanov:1990me}.  We can observe in Fig.~\ref{fig:allcurvatures-staro} how, for type I modes, the curvature perturbation is constant (the curves are superposed). This happens because those modes have exited the horizon during inflation. Therefore, the value of the curvature perturbation is frozen, thus remaining nearly constant even after they enter the particle horizon during the preheating stage (since when they enter the particle horizon, they also enter the RB)(See \eqref{eq:bardeen-super}). Furthermore, for type II modes, the curvature perturbation decays as $a^{-1}$ \eqref{eq:curvature-sub-horizon} (but as they enter the RB, they become a constant value). Once we have $\calR_k$ computed, we can obtain the density perturbations using \eqref{eq:density-pert-sup} and \eqref{eq:density-pert-sub}. Results are displayed in Fig. \ref{fig:alldensities-staro}.
\begin{figure}
    \centering
    \includegraphics[scale=0.35]{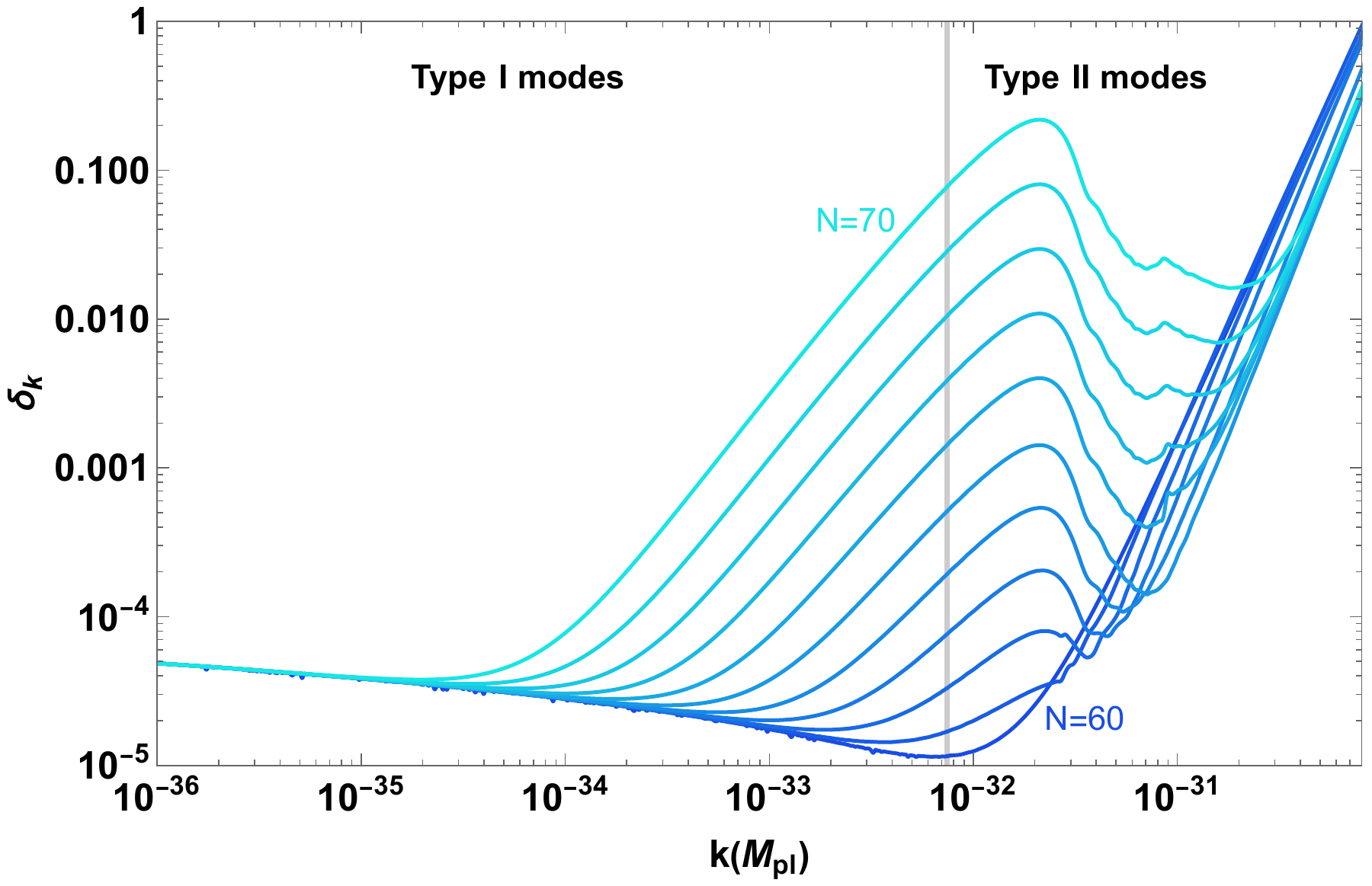}
    \caption{Averaged values of density perturbations $\delta_k$ as a function of the comoving wavenumber $k$ evaluated from $N=60$ (darker blue) to $N=70$ (lighter blue). The vertical grey line at $k_{\text{end}}$ separates the type I from the type II modes. This plot is made numerically up to $N=65$. Then, since at this point the potential \eqref{Staro} is well approximated by a quadratic one, we have used the parametric instability described in \cite{Martin:2019nuw} to extrapolate the evolution of the density perturbations by using the definition of resonance band. }
   % {Before $N=65$, we do not find it accurate to describe Starobinksy potential by a parabola.}}
    \label{fig:alldensities-staro}
\end{figure}

\begin{figure}
     \centering
     \begin{subfigure}[b]{0.49\textwidth}
         \centering         \includegraphics[width=\textwidth]{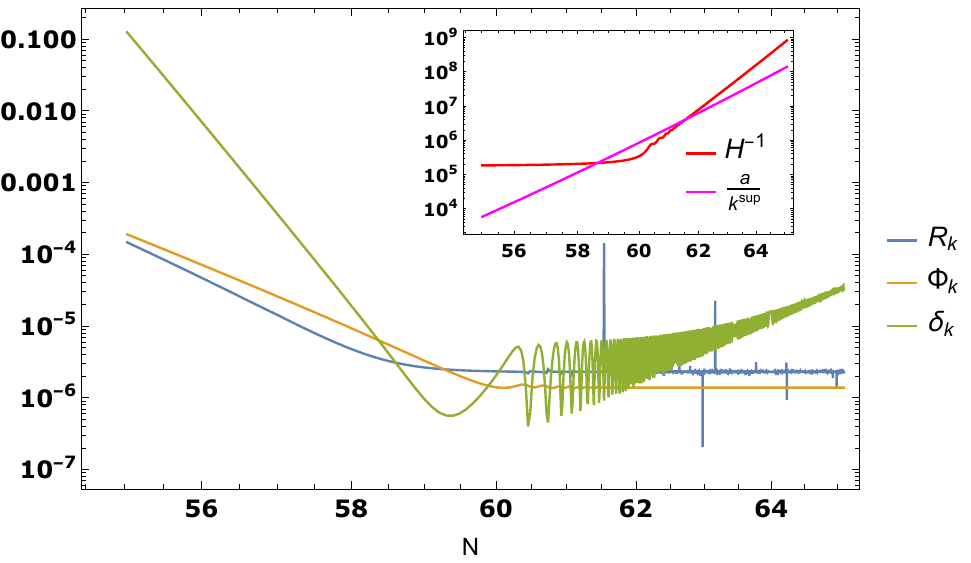}
         \caption{}
         \label{fig:evolution-sup}
     \end{subfigure}
     \hfill
     \begin{subfigure}[b]{0.49\textwidth}
         \centering         \includegraphics[width=\textwidth]{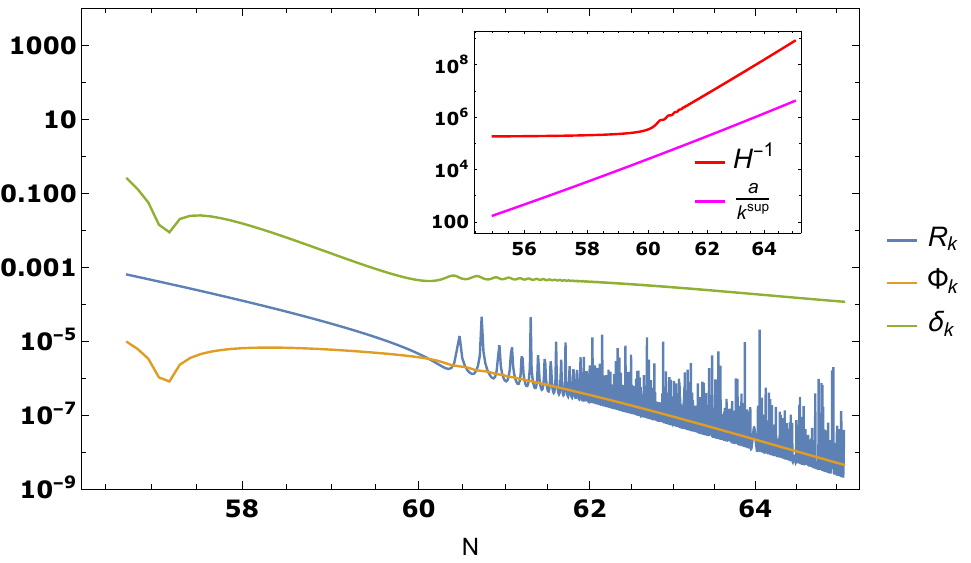}
         \caption{}
        \label{fig:evolution-sub}
     \end{subfigure}
        \caption{a) Evolution of $\calR_k$, $\Phi_k$ and $\delta_k$ during the end of inflation and preheating for a type I mode with comoving wavenumber $k^{I}=3\times10^{-33}\Mpl$. b) Evolution of $\calR_k$, $\Phi_k$ and $\delta_k$ during the end of inflation and preheating for a type II with comoving wavenumber $k^{II}=10^{-31}\Mpl$.}
        \label{fig:evolution-k}
\end{figure}

We observe that $\delta_k$ grows proportionately with the scale factor for type I modes as they enter the particle horizon, which agrees with \eqref{eq:density-pert-sup} (see also \eqref{eq:dominant-term}). As discussed before, the smaller $k$, the less time the mode will spend inside the horizon. Therefore, its amplification will be minimum, in contrast to the modes whose $k$ is close to the last scale to exit the horizon, $k_{\text{end}}$ (vertical grey line in Fig.~\ref{fig:alldensities-staro}). Those modes are sub-Hubble for more time, so their amplification will last longer. For type II modes, the smaller wave numbers also get amplified since the curvature perturbation is almost constant, as they are inside the resonance band during preheating. However, as $k$ increases, they remain approximately constant in time (since they are outside the resonance band), which agrees with \eqref{eq:density-pert-sub}. Fig.~\ref{fig:evolution-k} shows the behavior of $\calR_k$, $\Phi_k$ and $\delta_k$ for two particular modes: A type I mode with wavenumber $k^{I}=3\times10^{-33}\Mpl$, and a type II mode with $k^{II}=10^{-31}\Mpl$. For the former one (Fig.~\ref{fig:evolution-sup}), we see that both the curvature perturbation $\mathcal{R}_k$ and the perturbation $\Phi_k$ remain constant in contrast to $\delta_k$, which grows as $\sim a^{-1}$. On the other hand, for the type II mode (Fig.~\ref{fig:evolution-sub}), both $\mathcal{R}$ and $\Phi_k$ decays as $\sim a$ while the density perturbation $\delta_k$ remains almost constant.

%%%%%%%%%%%%%%%%%%%%%%%%%%%%%%%%%%%%

\section{Numerical characterization of PBH formation}\label{sec:results}

In this section, we give an account of PBH formation in the Starobinsky model, employing numerical calculations. First, we discuss the Press-Schechter (PS) formalism \cite{Press:1973iz} and, subsequently, compute the mass fraction. { We evaluate the threshold values of density fluctuations for each scale using the method proposed in \cite{Martin:2019nuw} i.e., whether the time the mode spends in the instability band is larger the the time needed to collapse and form PBH.} Then, for comparison, we compute the mass fraction using the Khlopov-Polnarev (KP) \cite{Khlopov:1980mg,Polnarev:1985btg,Khlopov:1982ef,Harada:2016mhb,Kokubu:2018fxy} formalism. Finally, we compute the PBH masses using the critical scaling method \cite{Maison:1995cc,Niemeyer:1999ak,Niemeyer:1997mt,Neilsen:1998qc,Musco:2012au,Snajdr:2005pr} for both formalisms.

%%%%%%%%%%%%%%%%%%%%%%%%%%%%%%%%%%%%%%%%%

\subsection{Press-Schechter formalism}\label{sec:PS}

Typically, to compute the fraction of collapsed objects into PBH, one uses the mass fraction $\beta_k$, which finds a precise definition in the Press-Schechter (PS) formalism \cite{Press:1973iz}. Under this formalism, we assume a Gaussian statistics $P$ for the density perturbations, $P(\delta)$, where the variance is typically given by the power spectrum of density perturbations, $\sigma^2(k)\simeq\calP_\delta(k)$. Then, the PBH mass fraction is expressed as \cite{Press:1973iz,Harada:2013epa}
\begin{equation}\label{eq:PS-formalism} \beta_k
= \frac{\dd\Omega_{\text{PBH}}(k)}{\dd \ln{M}}= 2 \int_{\delta_{c}}^{\delta_{\mathrm{max}}} P(\delta) \dd\delta 
\simeq\erfc\left[\frac{\delta_{c}}{\sqrt{2 \calP_\delta (k)}}\right]-\erfc\left[\frac{\delta_{\text{max}}}{\sqrt{2 \calP_\delta (k)}}\right],  
\end{equation}
where $\erfc$ is the complementary error function and $\delta_{\text{max}}$ is the maximum value for the density perturbation. We will take  $\delta_{\text{max}}=1$ to avoid the formation of PBH in the non-perturbative regime, where $\calP_{\delta}\gg1$ \cite{Banerjee:2022xft}. Therefore the enhanced perturbation modes will not contribute to increasing the mass fraction as we go to high values of $k$. The power spectrum for the density perturbations, $\calP_\delta(k)$, is defined as
\begin{equation}
    \calP_\delta(k)=\frac{k^3}{2\pi^2}|\delta_k|^2,
\end{equation}
{We evaluate the above quantity at the end of preheating for each mode.} As previously said, we will explore different preheating spans. The factor of 2 in the last equality comes from the Press-Schechter theory. 

Regarding the threshold values, as stated in the introduction, from the original analysis by Carr \cite{Carr:1975qj} (see also \cite{Escriva:2021,Sasaki:2018dmp,He:2019cdb,Yoo:2022mzl}), one can obtain an estimation given by $\delta_c\simeq w$, which is based on Jeans instability argument. The general idea is that the size of the over-density (considered as half the physical wavelength of the perturbation) at maximum expansion time must be more significant than the Jeans length, $R_J$, but also smaller than the particle horizon, $R_H=H^{-1}$ so that causality is preserved. There have been later refinements of this estimation, such as the one in \cite{Harada:2013epa}, where the authors obtained an analytical formula for the threshold as a function of $w$.
%\begin{equation}
%\delta_c \approx \frac{3}{5} \sin^2\LF \frac{\pi\sqrt{w} }{1+3w}\RF,
%\label{eq:delc}
%\end{equation}
In \cite{Escriva:2020tak} where the effect of the full shape of the compaction function during the collapse is considered. {However, all this is valid just for a perfect fluid formulation. Since our context is a scalar field, we will use the formulation given in \cite{Martin:2019nuw}, based on time assumptions, to estimate the threshold. In a nutshell, the time a perturbation $\delta_{\bm k}$ needs to collapse into a PBH is given by
\begin{equation}\label{eq:time-of-collapse}
    \Delta t_{\text{coll}}=\frac{\pi}{H[t_{\text{bc}}(k)]\delta_k^{3/2}[t_{\text{bc}}(k)]},
\end{equation}
where $t_\text{bc}(k)$ is the time at which each mode (type I or II) enters the instability band. Now, requiring that this time is equal to the time the mode is inside the instability band, $\Delta t_{\text{in}}=t_\text{r}-t_\text{bc}(k)$, where $t_r$ is the end of preheating, one can obtain an estimation for the threshold, the minimum value of $\delta_{\bm k}$ that can produce a PBH.}

%\tcr{Furthermore, we work here with approximating $w$ as a small non-zero constant for which \eqref{eq:delc} is applicable. However, the threshold $\delta_c$ is recently generalized for time-dependent $w$ case \cite{Escriva:2020tak,Papanikolaou:2022cvo} and we defer the corresponding study, of PBH during preheating, for future investigation.}
%We find it interesting to derive the expression for Jeans length at this point. 
%\tcr{In Newtonian gravity, the density contrast obeys the following equation in Fourier space \cite{Mukhanov:2005sc}
%\begin{equation}\label{eq:contrast_eq}
%    \ddot{\delta}_k+2H\dot{\delta}_k+(c_s^2k_{\text{p}}^2-\frac{\rho}{2\Mpl^2})\delta_k=0.
%\end{equation}
%Here, $c_s^2$ is the speed of sound of the fluid, which for a scalar field is defined as a function of the equation of state parameter through $c_s^2=w$ and $k_p=k/a$ is the physical wavenumber. This last is related to a physical wavelength by $\lambda_p=2\pi/k_p$.} 
{We are interested mainly in scales inside the horizon ($k>aH$) during the preheating phase. Since the effect of pressure is negligible, we can perform Newtonian perturbation theory, and the density contrast obeys the following differential equation \cite{Niemeyer:2019aqm,Reis:2003fs,Mukhanov:2005sc}
\begin{equation}\label{eq:density-contrast}
    \ddot{\delta}_k+2H\dot{\delta}_k+\left(c_s^2k_p^2-\frac{\rho}{2\Mpl^2}\right)\delta_k=0.
\end{equation}
Here, $c_s^2$ is the (effective) speed of sound, and $k_p=k/a$ is the physical wavenumber. This last is related to a physical wavelength by $\lambda_p=2\pi/k_p$. For a perfect fluid we have $c_s^2=w$ since $w$ is constant \cite{delaMacorra:2002du,Piattella:2013wpa}. However, for a scalar field (Appendix~\ref{Appw}), the equation of state $w$ is not constant in general. Therefore, the computation of the speed of sound is no longer trivial and must be done carefully. Following \cite{Cembranos:2015oya} (see also \cite{Hertzberg:2014iza}), the (effective) speed of sound for a general single-field case is
\begin{equation}
    c_s^2= \frac{\langle \delta p \rangle }{\langle \delta\rho \rangle} = \frac{\langle\frac{k^2}{a^2}\phi-V'(\phi)+V''(\phi)\phi\rangle}{\langle\frac{k^2}{a^2}\phi+3V'(\phi)+V''(\phi)\phi\rangle}.
\end{equation}
To obtain a simple expression for $c_s^2$, we will expand the Starobinsy potential around $\phi=0$ and up to fourth order, as done in Appendix~\ref{Appw}. Then, we average the field using \eqref{eq:field-param} and finally arrive at the expression
\begin{equation}\label{eq:speed-of-sound}
    c_s^2=\frac{\frac{k^2}{4M^2a^2}+\frac{\lambda}{4M^2}\langle \phi_0^2\rangle }{\frac{k^2}{4M^2a^2}+1+\frac{3\lambda}{8M^2}\langle \phi_0^2\rangle},
\end{equation}
we can see that in the limit of high $k$, the speed of sound reaches $c_s^2=1$. In Fig.~\ref{fig:speed-of-sound}, we depict the speed of sound as a function of $N$ for different values of $k$.
\begin{figure}[b]
    \centering         
    \includegraphics{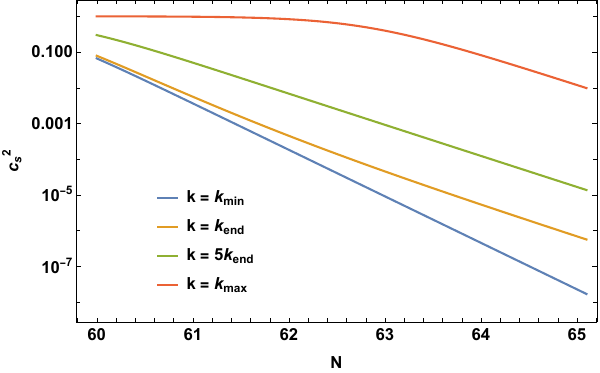}
    \caption{{The speed of sound as a function of $N$ for different $k$.}}
        \label{fig:speed-of-sound}
\end{figure}
In \eqref{eq:density-contrast} we see that for a physical wavelength greater that $\lambda_p>\lambda_J=\frac{\sqrt{8}\pi c_s}{\sqrt{\rho}}$ the perturbations will grow by gravitational collapse and perturbations with $\lambda_p<\lambda_J$ will develop acoustic oscillations \cite{Aviles:2011ak}. We consider Jeans length as half of the physical wavelength above which perturbations can collapse, i.e., $R_J=\frac{\lambda_J}{2}$, 
{Using eqn.~\eqref{eq:speed-of-sound} we find the following expression for Jeans length
\begin{equation}\label{eq:Jeans-Length}
    R_J=\left(\frac23\frac{\frac{k^2}{4M^2a^2}+\frac{\lambda}{4M^2}\langle \phi_0^2\rangle}{\frac{k^2}{4M^2a^2}+1+\frac{3\lambda}{8M^2}\langle \phi_0^2\rangle}\right)^{1/2}\pi R_H,
\end{equation}} 
where we have used $\rho=3H^2\Mpl^2$ {and $\lambda$ is given in \eqref{eq:lambda}}. Using this, we have that perturbations collapse if the following is satisfied
\begin{equation}\label{eq:jeansargument}
    R_J<\frac{\lambda_p}2<R_H \qquad\implies\qquad\frac{R_J}{\pi}<\frac{a}{k}<R_H.
\end{equation}
This condition means the only modes that can potentially collapse are those with physical wavelengths two times larger than the Jeans length, and this is what is called Jeans instability. This gives us a lower bound on the wavenumbers we must consider, which might collapse during the preheating.}
\begin{figure}
     \centering
     \begin{subfigure}[b]{0.49\textwidth}
         \centering         \includegraphics[width=\textwidth]{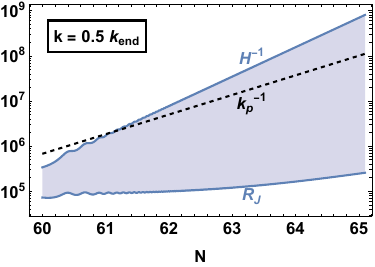}
         \caption{}
         \label{fig:jeans1}
     \end{subfigure}
     \hfill
     \begin{subfigure}[b]{0.49\textwidth}
         \centering         \includegraphics[width=\textwidth]{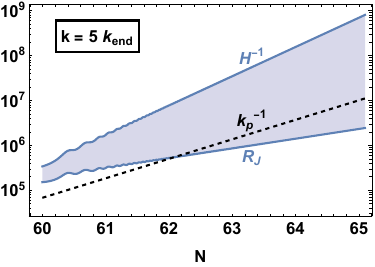}
         \caption{}
         \label{fig:jeans2}
     \end{subfigure}
        \caption{{Time evolution of Jeans length for two modes (left panel corresponds to a type I mode while the right panel corresponds to a type II mode.}}
        \label{fig:jeans-length}
\end{figure}
{ In  Fig.~\ref{fig:jeans-length}, we show the evolution of $R_J/\pi$ as a function of the number of e-folds for two different wavenumbers $k$ in the Starobinsky inflationary scenario. We observe that for modes with small $k$, Jeans length is unimportant since it is too small for them, and thus, as soon as they enter the horizon, they can potentially collapse. However, as we increase $k$, we see that it takes some time for the mode to grow and be able to collapse.
We can notice from  Fig.~\ref{fig:jeans-length} that the Jeans length grows at 
a smaller rate than $H^{-1}$. This makes some type II modes take some time before they grow enough and can collapse (see Fig.~\ref{fig:jeans-length}), which provides a physical mechanism to separate the non-linear regime induced by the smallest scales (high $k$) from our considerations.} 

\begin{figure}
    \centering
    \includegraphics{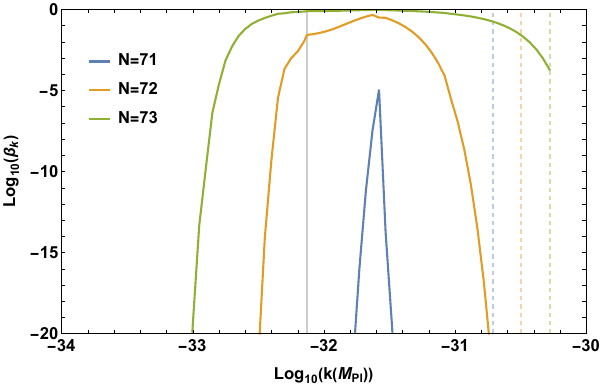}
    \caption{{Mass fraction of collapsed objects for different durations of preheating, computed using PS formalism. The threshold is obtained through time constraints with \eqref{eq:time-of-collapse}, and only modes that satisfy \eqref{eq:jeansargument} have been selected. Vertical dashed lines correspond to the smallest scale able to collapse at each evaluation step (with the same color code), computed using Jeans length argument.}}
    \label{fig:mass-fraction-PS-new}
\end{figure}

{Using the PS formalism from \eqref{eq:PS-formalism} we have computed in Fig.~\ref{fig:mass-fraction-PS-new} the mass fraction of collapsed objects. We have taken the threshold defined through time constraints using \eqref{eq:time-of-collapse}, and only modes that satisfy \eqref{eq:jeansargument} have been selected. Evaluation is made at different numbers of e-folds, from $N=71$ to $N=73$. As can be seen, as we increase the duration of preheating, the mass fraction increases since the modes have more time to collapse. Vertical dashed lines correspond to the smallest scale able to collapse for each duration of preheating and are computed using Jeans length argument. The vertical gray line marks the last scale to exit the horizon during inflation, $k_{\text{end}}$. We see that for $k<k_{\text{end}}$, we recover the results of \cite{Martin:2019nuw}. However, in our case, a shorter preheating is enough to obtain a similar mass fraction because of the contribution from type II modes.}

%%%%%%%%%%%%%%%%%%%%%%%%%%%%%%%%%%%%%%%%%%

\subsection{Khlopov-Polnarev formula}

\begin{figure}
    \centering
    \includegraphics{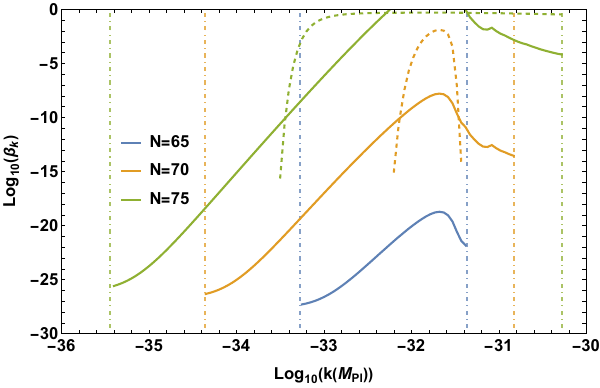}
    \caption{{KP formalism evaluated for different lengths of preheating. Vertical dotted-dashed lines correspond to the comoving Hubble radius (left) and smallest scale able to collapse (right), computed using Jeans length argument. The color code is the same as for the time steps.}}
    \label{fig:KP-mass-fraction}
\end{figure}

\begin{figure}
    \centering
    \includegraphics{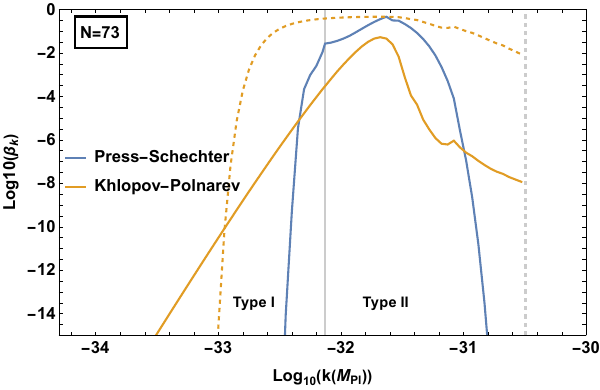}
    \caption{{Comparison between Press-Schechter formalism (blue) and Khlopov-Polnarev for small $\sigma(k)$ (orange continuous) and for high $\sigma(k)$ (orange dashed). The vertical grey line is at $k_{\text{end}}$, marking the limit between type I and II modes. The vertical grey dashed line is at the smallest scale and can collapse, following Jeans length argument.}}
    \label{fig:compare-PS-KP}
\end{figure}

Khlopov and Polnarev pioneered the study of PBH formation in a matter-dominated era \cite{Khlopov:1980mg,Polnarev:1985btg,Khlopov:1982ef}, where the effect of non-sphericity in the gravitational collapse plays a crucial role. They found that the fraction of collapsed objects could be expressed approximately as $\beta_k\simeq0.02\sigma^{5}(k)$ { for $\sigma(k)\ll1$}. This was refined by considering that if $\sigma(k)$ has a high value, the rise of density produces a pressure gradient that could prevent PBH formation. This added a factor of $\sigma^{3/2}(k)$ and therefore the mass fraction is now given by $\beta\simeq0.02\sigma^{13/2}(k)$. Later on \cite{Harada:2016mhb} refined this criterion, obtaining the following semi-analytic formula
\begin{equation} \label{eq:KP-formula}
    \beta_k{(\sigma<0.01)}\simeq0.05556\,\sigma^{5}(k),
\end{equation}
where the effect of the increasing pressure gradient is not considered since it heavily relies on the matter model. This formula \eqref{eq:KP-formula} agrees with numerical simulations. However, {for $0.01<\sigma(k)<\mathcal{O}(1)$,} the numerical production rate seems to be higher than this power-law formula (see Fig.1 of \cite{Harada:2016mhb}) {and thus \eqref{eq:KP-formula} must be replaced by  \cite{Kokubu:2018fxy}
\begin{equation}\label{eq:KP-high-sigma}
    \beta_k(\sigma>0.01)\simeq\frac12\left[1-\text{Erf}\left(\frac{0.11}{\sqrt{2}\sigma(k)}\right)\right].
\end{equation}}
{Therefore, we use \eqref{eq:KP-formula} and \eqref{eq:KP-high-sigma} to compute the mass fraction for the values of $\sigma(k) \lessgtr 0.01 $ and $\sigma(k)<1$, respectively. The estimates are depicted in Fig.~\ref{fig:KP-mass-fraction}, where we plot the estimates from KP formalism using both approaches \eqref{eq:KP-formula} (continuous) and \eqref{eq:KP-high-sigma} (dahsed). 
%These results are in the context of shorter duration of preheating (5 e-folds). Beyond  $N=65$, there is no appreciable contribution from \eqref{eq:KP-high-sigma}.  %However, this is not the case for $N=70$ and $N=75$, where We have noticed that $\sigma>\mathcal{O}(1)$ $\beta_k\simeq0.5$. 
{We can notice in Fig.~\ref{fig:KP-mass-fraction} that the KP formalism is very sensitive to the preheating number of e-folds. The mass fraction drastically increases as we increase the number of e-folds during the preheating. }
It is worth noting that the KP formula only applies to the exact matter-dominated era, and the preheating cannot be strictly taken as that. However, given the smallness of the effective equation of state during preheating (see Appendix~\ref{Appw}), we get estimates subjected to negligible effects of non-zero pressure, providing reasonable bounds of PBH formation using the KP formalism. We can see {in Fig.~\ref{fig:compare-PS-KP}} that for lower values of $\sigma(k)$ the KP formula {\eqref{eq:KP-formula} (blue)} gives a higher estimate than the PS formalism {\eqref{eq:PS-formalism} (orange continuous)}. However, as $\sigma(k)$ increases, this last has a higher estimate for the PBH abundance, {unless the expression for high $\sigma(k)$, eqn.~\eqref{eq:KP-high-sigma} is used (orange dashed)}.}
%\begin{figure}
  %   \centering
  %   \begin{subfigure}[b]{0.49\textwidth}
  %       \centering         \includegraphics[width=\textwidth]{Images/mass-fraction-5efolds.pdf}
  %       \caption{}
  %       \label{fig:KP-5}
 %    \end{subfigure}
 %    \hfill
   %  \begin{subfigure}[b]{0.49\textwidth}
   %      \centering         \includegraphics[width=\textwidth]{Images/mass-fraction-10efolds.pdf}
   %      \caption{}
  %       \label{fig:KP-10}
  %   \end{subfigure}
  %      \caption{\tcr{Mass fraction of collapsed objects evaluated at different points: a) $N=65$ e-folds and b) $N=70$ e-folds as a function of the comoving wavenumber $k$. Inflation ends at $N=60$. For each case, it is shown both the Press-Schechter (blue \tcr{tones}) and Khlopov-Polnarev (red) formalisms. Threshold value are taken to be $\delta_c=10^{-4},5\times10^{-4},10^{-3}$ for the Press-Schechter formalism. The vertical grey line at $k_{\text{end}}$ separates the type I from the type II modes. On each plot, vertical dashed lines mark the comoving Jeans length (for each value of $w\simeq\delta_c$) and the left vertical gray dashed line marks the comoving Hubble radius.}}
%        \label{fig:KP}
%\end{figure}

%%%%%%%%%%%%%%%%%%%%%%%%%%%%%%%%%%%%%%%

\subsection{PBH associated mass}

Let us turn to determining the mass associated with these PBH. For $\delta_k\geq\delta_c$, and $\delta_k\simeq\delta_c$, the PBH mass follows a scaling relation with $\delta_k$ given by
\begin{equation}\label{eq:scaling}
    M_{\text{PBH}}(k)=M_{\text{H}}\kappa(\delta_k-\delta_c)^{\gamma},
\end{equation}
where $\kappa,\gamma$ are constants ($\gamma$ being dependent on the equation of state parameter  $w$), and $M_{\text{H}}$ is the horizon mass at the time of PBH formation, which is given by
\begin{equation}\label{eq:horizon-mass}
    M_{\text{H}}= \frac{4\pi}{3} \rho H^{-3}\simeq \frac{4\pi}{3} \rho_{end} {a_{end}^3}\left(\frac{1}{aH}\right)^3\, .
\end{equation}
We have used the fact that the energy density, during matter-domination, decays as
\begin{equation}    \rho\simeq\rho_{\text{end}}\left(\frac{a}{a_{\text{end}}}\right)^{-3}.
\end{equation}
References \cite{Maison:1995cc,Niemeyer:1999ak,Niemeyer:1997mt,Neilsen:1998qc,Musco:2012au,Snajdr:2005pr} present a vast account of numerical studies where \eqref{eq:scaling} was derived. Parameter $\kappa$ ranges from $\kappa\simeq2.4$ to $\kappa\simeq12$ \cite{Niemeyer:1999ak}. We will consider $\kappa=4$ to estimate the mass. Regarding the exponent $\gamma$, the scenario where $w=0$ corresponds to a singular point, in the sense stated in \cite{Maison:1995cc}. However, in the present case, $\gamma$ seems to approach a non-vanishing value when $w\rightarrow0$. This situation was first analyzed in \cite{Snajdr:2005pr}, where a $\gamma=0.1057$ was obtained. We will use this value in our numerical evaluation of the PBH-associated mass, {accounting the smallness of our effective equation of state $w_{eff}$ (see App.~\ref{Appw})}.

\begin{figure}
     \centering
     \begin{subfigure}[b]{0.49\textwidth}
         \centering         \includegraphics[width=\textwidth]{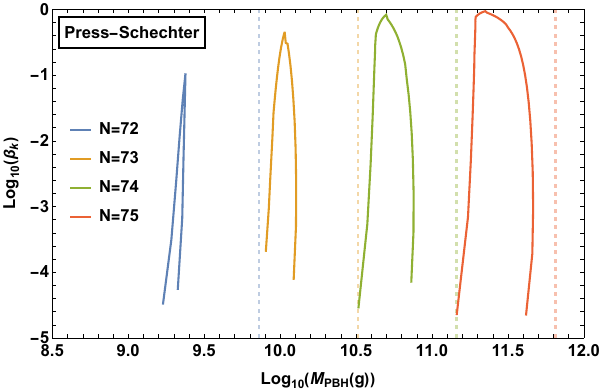}
         \caption{}
         \label{fig:mass-fraction-grams-PS}
     \end{subfigure}
     \hfill
     \begin{subfigure}[b]{0.49\textwidth}
         \centering         \includegraphics[width=\textwidth]{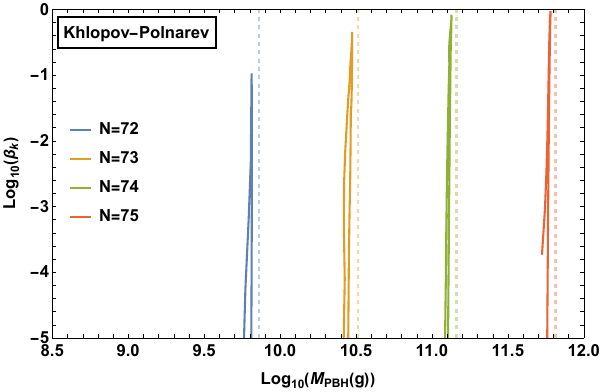}
         \caption{}
         \label{fig:mass-fraction-grams-KP}
     \end{subfigure}
        \caption{{Mass fraction of collapsed objects evaluated at different points after inflation for the two estimations under consideration, Press-Schechter \eqref{eq:PS-formalism} (a) and Khlopov-Polnarev \eqref{eq:KP-high-sigma} (b). Thresholds are obtained by imposing time constraints by \eqref{eq:time-of-collapse}, and only modes that satisfy \eqref{eq:jeansargument} have been selected. The vertical dotted lines represent the horizon mass at the evaluation time, following the same color code of the mass fractions.}}
        \label{fig:mass-fraction-grams-new}
\end{figure}

{In Fig.~\ref{fig:mass-fraction-grams-new}, we represent the mass fraction as a function of the PBH mass, which is obtained using \eqref{eq:scaling} for the PS and KP frameworks. The modes selected to produce this plot have to fulfill two requirements. First, only modes with $\delta_k\gtrsim \delta_c$ are considered so that \eqref{eq:scaling} can be used. Second, we have to keep up with the requirement of Jeans instability, which means satisfying the inequality \eqref{eq:jeansargument}. The mass fraction has a peak centered close to the horizon mass at each evaluation step. 
One can notice in Fig.~\ref{fig:mass-fraction-grams-new}, the mass fraction over $M_{\rm PBH}$ is somewhat broadly distributed with PS, compared with KP formalism. From Fig.~\ref{fig:compare-PS-KP} and Fig.~\ref{fig:mass-fraction-grams-new}, we can learn how modes spanning a range of $k$ contribute to the formation of different PBH mass ranges and how this depends mainly on the duration of preheating. 
As we expand the preheating duration, the PBH' mass increases since horizon mass grows in time; see \eqref{eq:horizon-mass}. However, the shape of $\beta_k$ does not seem to change. Finally, the effect of the threshold is not very important for PBH formation with masses close to the horizon but tends to increase its effect when we consider smaller PBH. We conclude that the PBH formation is higher than the previous estimates
\cite{Martin:2019nuw} by more than 5 orders of magnitude even with a reduced preheating duration. This is because of the additional non-negligible contribution from the type II modes, and it is also somewhat expected in \cite{Martin:2019nuw}. Thus, our study brings a quantitative estimate of all modes that can potentially collapse and form PBH with preheating instabilities.  }
%\begin{figure}
%     \centering
%     \begin{subfigure}[b]{0.49\textwidth}
%         \centering         \includegraphics[width=\textwidth]{Images/mass-fraction-5efolds-grams.pdf}
 %        \caption{}
  %       \label{fig:mass-fraction-5-grams}
  %   \end{subfigure}
  %   \hfill
  %   \begin{subfigure}[b]{0.49\textwidth}
  %       \centering         \includegraphics[width=\textwidth]{Images/mass-fraction-10efolds-grams.pdf}
  %       \caption{}
 %        \label{fig:mass-fraction-10-grams}
  %   \end{subfigure}
  %      \caption{\tcr{Mass fraction of collapsed objects evaluated at different points after inflation: a) 5 e-folds and b) 10 e-folds for the two models under consideration, Press-Schechter (blue tones) and Khlopov-Polnarev (red tones). Threshold values are taken to be $\delta_c=10^{-4},5\times10^{-4},10^{-3}$ from darker to lighter blue/red, respectively. The vertical grey line marks the horizon mass at the evaluation time.}}
%        \label{fig:mass-fraction-grams}
%\end{figure}

%%%%%%%%%%%%%%%%%%%%%%%%%%%%%%%%%%%%%%%%%

\section{Conclusions}\label{sec:conclusions}

The possibility of PBH formation during the inflationary preheating precedes the later stages of radiation and matter-dominated eras. Since any inflationary model with a graceful exit to reheating constitutes the preheating stage, it is vital to understand the details of the possible collapse of these primordial modes. Even though these PBH, which could potentially get formed during preheating, are rather not large enough to survive until now to act as dark matter, they would most likely contribute to the large PBH at a later time due to the primordial clustering \cite{DeLuca:2020jug,DeLuca:2021hcf,DeLuca:2022bjs}. Given this, we revisit the PBH formation criteria during the preheating stage and {extend the study made previously in \cite{Martin:2019nuw} by also considering the type II modes}. Furthermore, contrary to the usual assumption to discard any pressure effects on PBH formation during preheating, we carefully consider the effects of small non-zero pressure and explore the consequences of the collapse of primordial modes. 
To carry out this program, we have numerically computed the evolution of curvature and density perturbations during inflation and the subsequent oscillatory period of preheating. We worked in the framework of Starobinsky inflation, which is the most realistic in the context of recent observational consistency of the model \cite{Planck:2018jri}. Therefore, our preheating PBH study contains results that extend the study performed in the context of the chaotic inflationary model, which is ruled out by observations \cite{Martin:2019nuw}. We categorize the primordial modes as type I and type II, depending on whether the modes exit the horizon during inflation. Type I modes are the ones mainly considered in earlier studies \cite{Martin:2019nuw}, corresponding to those that exit the horizon during inflation and enter the particle horizon later in the preheating stage and fall in the window of RB (See Fig.~\ref{fig:parametric-instability} and \eqref{eq:RB}).  However, in this paper, we further studied type I modes in the context of Jeans instability. Type II modes, a new element of our investigation, remain sub-horizon during inflation but quickly evolve to create over-densities in the preheating stage. Among these type II modes, a sub-class of them that enter the RB (c.f. Fig.~\ref{fig:scales} and \ref{eq:RB}) evolves very similarly to type I, as their density perturbation grows linearly with the scale factor. The remaining type II modes that do not fall under RB have their density perturbation almost constant (See \eqref{eq:density-pert-sub}). By considering type II modes, one can achieve faster formation of PBH. Since these modes grow faster, they need less time to collapse. 
We have shown that in the context of relatively short periods of preheating (i.e., 1-5 e-folds after the end of inflation), type II modes are more dominant ones to experience Jeans instability and eventually collapse to form PBH, compared to type I modes. In the context of more extended periods of preheating (i.e., around 10 e-foldings after the end of inflation), we found that both type I and type II modes comparably contribute to the mass fraction of PBH which can be seen in Fig.~\ref{fig:alldensities-staro}. 
To compute the mass fraction of collapsed objects, $\beta_k$, we have used the Press-Schechter formalism \eqref{eq:PS-formalism} \cite{Press:1973iz,Harada:2013epa} and Khlopov-Polnarev formulae \eqref{eq:KP-formula} and \eqref{eq:KP-high-sigma} \cite{Harada:2016mhb,Khlopov:1980mg,Polnarev:1985btg,Khlopov:1982ef}, which suits best to estimate the collapse structures during the inflaton matter-dominated universe. Results are shown in {Figs.~\ref{fig:mass-fraction-PS-new} and \ref{fig:KP-mass-fraction}}, where both formalisms are compared. In the case of applying PS formalism in our study, {we have calculated the threshold of density contrast for each mode by equating the duration of the collapse with the time each perturbed mode resides in the instability band. Further, we used effective sound speed for scalar field fluctuations to evaluate Jeans length.} Our results concerning KP formalism are limited to the assumption of an exact matter-dominated phase.   {Since, in our case, the effective equation of state during the preheating is nearly zero, we can neglect the effects of pressure. However,} We defer detailed studies involving the role of non-zero pressure as a subject of future investigation. One can see that, for Press-Schechter formalism, the formation of PBH is enhanced compared to the Khlopov-Polnarev case when the density contrast is high {unless we take into account \eqref{eq:KP-high-sigma}}. With a low-density contrast, the semi-analytical formula~\eqref{eq:KP-formula} provides a better fit than the numerical solution. Finally, {Fig.~\ref{fig:mass-fraction-grams-new}} shows the mass fraction as a function of the mass of the PBH formed for both formalisms. One can see that as the preheating goes on, more and more PBH form with higher probability and mass since they form close to the critical point where $\delta_k\simeq\delta_c$. {It is interesting that both PS and KP formulations give comparable estimates as we increase the number of e-folds of the preheating. KP formalism, in particular, is very sensitive to the preheating duration, which is expected because of the way mass fraction $\beta_k$ scales according to $\sigma(k)$ in \eqref{eq:KP-formula} and \eqref{eq:KP-high-sigma}.}
In any case, the peak of the mass fraction is constrained by the horizon mass at the time of formation. Our study opens new doors for investigating PBH formation during the preheating stage. We explored the wide range of density perturbations that experience Jeans instability and collapse and addressed the problem in Starobinsky inflation. In future work, we aim to expand our study and methodology to more general models like $\alpha$-attractors \cite{alphaatr}. Moreover, our study gives more precise estimates for the mass fraction of PBH during the inflaton-like matter-dominated era, and it is essential to extend our analysis in the scope of primordial clustering of PBH to decipher the PBH's role as dark matter now fully. Also, the PBH of masses {$10^{9}g$ to $10^{12}g$} we predict {(see Fig.~\ref{fig:mass-fraction-grams-new})}, in the context of different durations for the preheating period, could act as primordial seeds for the supermassive black holes we see today \cite{DeLuca:2022bjs,Bean:2002kx,DeLuca:2021pls,DeLuca:2020jug}. 

%%%%%%%%%%%%%%%%%%%%%%%%%%%%%%%%%%%%%%%%

\appendix

\section{{Effective equation of state}}
\label{Appw}

\begin{figure}
    \centering
    \includegraphics[scale=1]{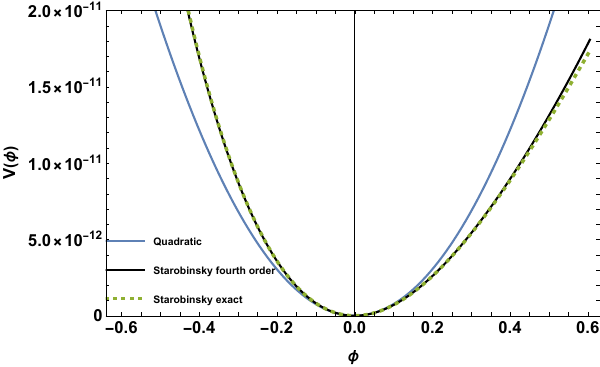}
    \caption{{Comparison between quadratic inflation ($V=\frac{M^2}{2}\phi^2$) in blue, the series expansion \eqref{eq:staro-pot-expansion} of the Starobinsky potential in black and the full version of it (eqn.~\eqref{Staro}) in dashed green.}}
    \label{fig:potentials}
\end{figure}

In this section, we give an analytical parametrization of the equation of state. We call this the effective equation of state, $w_{\text{eff}}$. The main reason to do this is because the universe does not behave as purely matter-dominated just right after the end of inflation. There is some transition period from inflation into matter-domination. In \cite{Garcia:2020wiy}, the effective equation of state was derived using a potential of the form $V(\phi)=\frac{m^2}{2}\phi^n$, just after the end of inflation, giving 
\begin{equation}
    w_{\text{eff}}=\frac{n-2}{n+2}.
\end{equation}
For $n=2$, we immediately see that $w_{\text{eff}}=0$, and thus, the universe is perfectly approximated by a matter-dominated one after inflation. However, as we will see, this is not the case for the Starobinsky model. Let us consider, for example, the potential \eqref{Staro}. By making a series expansion around $\phi=0$ up to the fourth order, we obtain
\begin{equation}\label{eq:staro-pot-expansion}
    V(\phi)\simeq\frac{M^2}2\phi^2+\frac{\lambda_3}{3}\phi^3+\frac{\lambda}4\phi^4,
\end{equation}
where $M$ is the scalaron mass, to be normalized with CMB observations, and the parameters $\lambda_3$ and $\lambda$ are defined as
\begin{equation}\label{eq:lambda}
    \lambda_3=-\sqrt{\frac{3}{2}}M^2\qquad\lambda=\frac{7}{9}M^2.
\end{equation}
Both of these parameters cause the potential to be different from the purely quadratic one (see Fig.~\ref{fig:potentials} for illustration). 
%With this motivation in mind, we derive, following \cite{Garcia:2020wiy}, an analytical expression of $w$ for a generic potential of the form \eqref{eq:staro-pot-expansion}.}
Multiplying the Klein-Gordon equation \eqref{KG} by $\phi$ and averaging over one period of oscillation, we obtain the following for the potential \eqref{eq:staro-pot-expansion} 
\begin{equation}
    \langle\phi V'(\phi)\rangle\simeq M^2\langle\phi^2\rangle+\lambda_3 \langle\phi^3\rangle+ \lambda\langle\phi^4\rangle,
\end{equation}
where we have applied the virial theorem $\langle\phi V'(\phi)\rangle\simeq\langle\dot{\phi}^2\rangle$. Using this, the background energy density can be written as
\begin{equation}
    \langle \rho \rangle \simeq\frac{\langle\dot\phi^2\rangle}{2}+\langle V(\phi)\rangle\simeq M^2 \langle\phi^2\rangle+\frac{5}{6}\lambda_3\langle\phi^3\rangle+\frac{3}{4}\lambda\langle\phi^4\rangle,
\end{equation}
and the background pressure as
\begin{equation}
    \langle P\rangle \simeq\frac{\langle{\dot\phi}^2\rangle}{2}-\langle V(\phi)\rangle\simeq \frac{\lambda_3}{6}\langle\phi^3\rangle+ \frac{\lambda}{4}\langle\phi^4\rangle.
\end{equation}
Now, the evolution of the field $\phi$ is parameterized by \cite{Garcia:2020wiy}
\begin{equation}\label{eq:field-param}
    \phi(t)\simeq\phi_0(t)T(t),
\end{equation}
where $\phi_0(t)=\phi_{\text{end}}\left(\frac{a_{\text{end}}}{a}\right)^{3/2}$ encodes the decaying amplitude of the field due to the redshift of the universe and $T(t)$ is an oscillatory periodic asymmetric function (due to the $\lambda_3$ coefficient). Its average value can be computed as $\langle T(t)^n\rangle\simeq\frac{2}{n+2}$ for $n$ even. Following \cite{Cembranos:2015oya}, for $n$ odd, the sinusoidal resulting function oscillates around zero and is suppressed by the averaging. Thus, we will only consider the average of even powers. Using the averaging of $\langle T(t)^n\rangle$ and \eqref{eq:field-param} into the background energy and pressure we get
\begin{equation}
    \begin{split}
        \langle \rho\rangle &\simeq \langle V(\phi_0) \rangle \\
        \langle P\rangle &\simeq \frac{\lambda}{12}\langle\phi_0^4\rangle.
    \end{split}
\end{equation}
Now, the effective equation of state can be computed as
\begin{equation}\label{eq:EOS-averaged}
    w_{\text{eff}}=\frac{\langle P\rangle}{\langle \rho \rangle}\simeq \frac{\lambda}{12}\frac{\langle\phi_0^4\rangle}{\langle V(\phi_0) \rangle} =\frac{\frac{\lambda}{6M^2}\langle \phi_0^2\rangle}{1+\frac{\lambda}{2M^2} \langle\phi_0^2\rangle}.
\end{equation}
Here, we can see that after the end of inflation, the effective equation of state is not exactly zero. It starts with small positive values and approaches zero as preheating continues, reaching the approximated matter-dominated stage. In Fig.~\ref{fig:equation-of-state}, we observe the effect of these extra terms in the expansion of the Starobinsky potential, where the effective version is compared with the numerical one, obtained from \eqref{eq:EOS-full}.

%%%%%%%%%%%%%%%%%%%%%%%%%%%%%%%%%%%%%%%

\section{Initial conditions for inflation}\label{appA}

This appendix provides a general procedure to obtain initial conditions for a given single-field model with potential $V=V(\phi)$. It is based on the slow-roll approximation, and therefore, the resulting equations are only valid for small slow-roll parameters, that is, at the beginning of the inflationary era. Let us start with Friedmann equations written in cosmic time:
\begin{equation}\label{eq:first-FR}
H^2=\frac1{3\Mpl^2}\left(\frac{\dot{\phi}^2}2+V\right),
\end{equation}
\begin{equation}\label{eq:second-FR}
\dot{H}+H^2=-\frac1{3\Mpl^2}\left(\dot{\phi}^2-V\right).
\end{equation}
Substituting \eqref{eq:first-FR} into \eqref{eq:second-FR}, we obtain the following relation for the derivative of the Hubble factor
\begin{equation}\label{eq:relation}
    \dot{H}=-\frac{\dot{\phi}^2}{2\Mpl^2}.
\end{equation}
Now, the first slow-roll parameter $\epsilon$ is defined in terms of the Hubble factor as:
\begin{equation}\label{eq:slow-roll-param}
    \epsilon=-\frac{\dot{H}}{H^2}.
\end{equation}
Using \eqref{eq:relation} together with \eqref{eq:slow-roll-param} we have
\begin{equation}\label{eq:relation2}
    \frac{H}{\dot{\phi}}=\frac1{\sqrt{2\epsilon}\Mpl}.
\end{equation}
The number of e-folds can be expressed as $\dd N=H\dd t=\frac{H}{\dot{\phi}}\dd\phi$. Thus, substituting this into \eqref{eq:relation2} we have
\begin{equation}\label{eq:relation3}
    \dd N=\frac{1}{\sqrt{2\epsilon}\Mpl}\dd\phi.
\end{equation}
Another common way of defining the first slow-roll parameter regarding the potential is there. We call it $\epsilon_V$, and it is computed as
\begin{equation}\label{eq:slow-roll-potential}
    \epsilon_V=\frac{\Mpl^2}{2}\left(\frac{V_{,\phi}}{V}\right)^2,
\end{equation}
where the subscript ``$_{,\phi}$'' refers to derivation with respect to the field $\phi.$. During inflation, the slow-roll parameters can be approximated, $\epsilon\simeq\epsilon_V$, so substituting \eqref{eq:slow-roll-potential} into \eqref{eq:relation3} gives us a relation between the field and the number of e-folds, that is:
\begin{equation}     
N=-\frac1{\Mpl^2} \int_\phi^0 \left(\frac{V}{V_{,\phi'}}\right)\dd\phi'.
\end{equation}
Usually, the number of e-folds is counted backward; $N=60$ marks the beginning of inflation, and $N=0$ the end. The number of e-folds in our code is counted forward, but this is just for numerical purposes. To obtain $\epsilon(N)$, we have to go back to \eqref{eq:relation2} and use the relation $\dd N=H\dd t$ again to transform the derivative with respect to $t$ to a derivative with respect to $N$. Then, using \eqref{eq:relation3} we obtain
\begin{equation}
    \epsilon(N)=\frac{1}{2\Mpl^2}\left(\frac{\dd\phi}{\dd N}\right)^2=\frac12\left(\frac{V_{,\phi}}{V}\right)^2.
\end{equation}
Here, in the last step, we have used \eqref{eq:slow-roll-potential} as an additional way to compute $\epsilon(N)$. Depending on the model, one way will be better than the other. Let us finally see how to obtain $\frac{\dd\phi}{\dd N}$. Using the Klein-Gordon equation \eqref{KG}, neglecting $\ddot{\phi}$ and transforming $\dot{\phi}$ into a derivative with respect to $N$ we have
\begin{equation}
    \frac{\dd\phi}{\dd N}=-\frac{V_{,\phi}}{3H^2}.
\end{equation}
Now, using \eqref{eq:first-FR} to substitute $H^2$ and using the slow-roll approximation ($V\gg\dot{\phi}^2$) we obtain:
\begin{equation}
    \frac{\dd\phi}{\dd N}=-\Mpl^2\frac{V_{,\phi}}{V}.
\end{equation}

%%%%%%%%%%%%%%%%%%%%%%%%%%%%%%%%%%%%%%%%

\section{Initial conditions for perturbations}\label{appB}

\begin{figure}
     \centering
     \begin{subfigure}[b]{0.32\textwidth}
         \centering         \includegraphics[width=\textwidth]{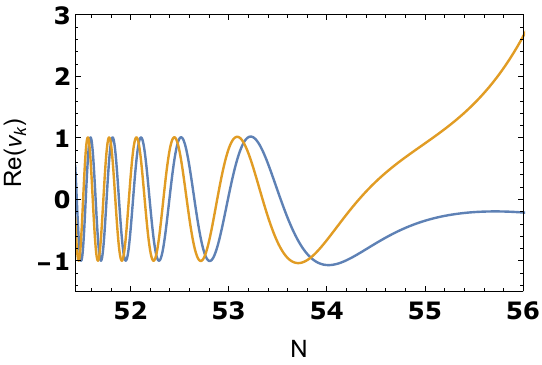}
         \caption{}
         \label{fig:muk-real}
     \end{subfigure}
     \hfill
     \begin{subfigure}[b]{0.32\textwidth}
         \centering         \includegraphics[width=\textwidth]{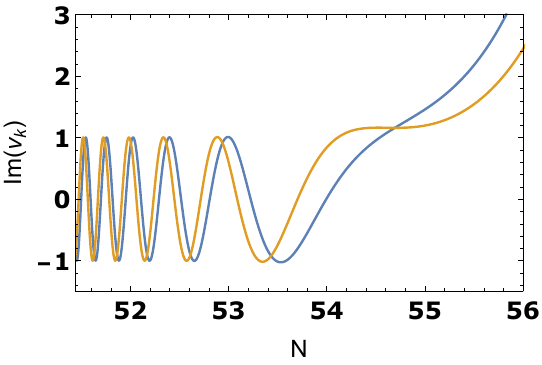}
         \caption{}
         \label{fig:muk-imag}
     \end{subfigure}
     \hfill
     \begin{subfigure}[b]{0.32\textwidth}
         \centering         \includegraphics[width=\textwidth]{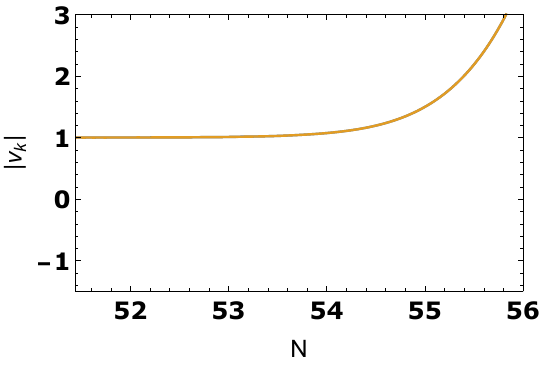}
         \caption{}
         \label{fig:muk-abs}
     \end{subfigure}
        \caption{Evolution of the Mukhanov variable for two different choices of conformal time and for $k=10^{-34}\Mpl$. Hubble-crossing occurs at approximately 54 e-folds. In blue, conformal time is set to zero at the end of inflation $\eta(t_{end})=0$ and in orange $\eta(t_{end})=\frac{\pi}{4}$. In a) the real part is plotted, in b) the imaginary part, and in c) the modulus $|v_k|$. In this last plot, both plots are superimposed, showing that the choice of conformal time does not affect $|v_k|$, but indeed, it affects the real and imaginary parts. $v_k$ is rescaled to have unitary amplitude while sub-Hubble.}
        \label{fig:muk-variable}
\end{figure}

{This section explains how we choose the initial conditions to solve \eqref{MScosmic}. We impose the initial conditions for all the fluctuations during inflation when the modes are deep inside the horizon $k \gg aH$. 
The Mukhanov-Sasaki variable is usually expanded as 
\begin{equation}
    \hat{v}(\eta, \textbf{x}) = \int \frac{d^3k}{\LF 2\pi \RF^{3/2} } \Bigg[ a_\textbf{k} v_k e^{i\textbf{k}\cdot \textbf{x}} +a^\dagger_\textbf{k} v^\ast_k e^{-i\textbf{k}\cdot \textbf{x}}  \Bigg]
\end{equation}
where $a_\textbf{k},\, a_\textbf{k}^\dagger$ are the creation and annihilation operators that satisfy the canonical commutation relations. The general solution of the Mukhanov-Sasaki equation \eqref{MSconformal} during the inflationary de Sitter phase is 
\begin{equation}
    v_k(\eta) = \frac{A_k}{\sqrt{2k}} \LF 1-\frac{i}{k\eta} \RF e^{-ik\eta} + \frac{B_k}{\sqrt{2k}} \LF 1+\frac{i}{k\eta} \RF e^{ik\eta}
\end{equation}
where $A_k,\, B_k$ are the Bogoliubov coefficients. 
The canonical commutation relation of the MS field leads to \cite{Mukhanov:1990me} 
\begin{equation}
    v_kv_k^{\ast\prime} -v_k^\ast v_k^{\prime} = i 
\end{equation}
which yields conditions on the coefficients 
\begin{equation}
    \vert A_k \vert^2-\vert B_k\vert^2 =1
\end{equation}
The Bunch-Davies vacuum is 
\begin{equation}
    A_k = 1,\quad B_k =0\,. 
\end{equation}
The Bunch-Davies vacuum corresponds to the positive energy state when the mode is sub-horizon $k\gg aH$ 
\begin{equation}
    v_k\Bigg\vert_{k\gg aH} = \frac{1}{\sqrt{2k}} e^{-ik\eta}
    \label{modef}
\end{equation}}
  The \eqref{modef} can be written in a trigonometric form as:
\begin{equation}\label{eq:BD-trig}
    v_k(\eta)=\frac{1}{\sqrt{2k}}\left[\cos{(k\eta)}-i\sin{(k\eta)}\right].
\end{equation}
This complex oscillatory function has a constant amplitude given by $1/\sqrt{2k}$. Let us define the following two conditions:
\begin{itemize}\setlength\itemsep{-0.3em}
    \item[i)] Real and imaginary parts must be synchronized, in the sense that $$\cos{(k\eta)}^2+\sin{(k\eta)}^2=1,$$ so that the amplitude does not get affected.
    \item[ii)] The initial computational time must be such that we let the mode evolve for a sufficient number of e-folds before the freezing (Hubble-crossing) point is reached.
\end{itemize}
As long as these two conditions apply, then the choice of initial computational time (conformal or cosmic) should have no impact on the final result for the modulus of Mukhanov's variable $|v_k|$ (see Fig. \ref{fig:muk-variable} for details). This is because the freezing point acts as an attractor and always occurs at the same time and with the same amplitude, provided conditions i) and ii) are satisfied and that we refer to $|v_k|$. We will choose the following simple initial condition\footnote{In essence, this is like setting $t_i$ such that $k\eta(t_i)=2n\pi$, with $n\in\mathbb{N}$. This fully satisfies condition i). We can always do that since conformal time is defined up to a constant and therefore is rather arbitrary \cite{Habib:2005mh}.\label{footnote}} for all modes:
\begin{equation}\label{eq:BD-init1}
    v_k(t_i)=\frac1{\sqrt{2k}}.
\end{equation}
This, of course, satisfies condition i) (see footnote \ref{footnote}) and, if $t_i$ is such that it lets the mode evolve for some e-folds, then also condition ii). Following \cite{Ballesteros:2017fsr,Wolfson:2021utn}, the starting time for the computations can be set for each mode to just a few e-folds before Hubble crossing, usually 2 or 3. This is because, at this point, the mode is still well described by the Bunch-Davies vacuum. Also, in doing so, we avoid the computation of unnecessary oscillations that the mode does. This consumes memory, which translates into more waiting time.

Let us see now how to obtain the initial condition for $\dot{v}_k$. Using the definition of conformal time we have
\begin{equation}\label{eq:deriv-conf}
    \dd\eta=\frac{\dd t}{a(t)}\qquad\rightarrow\qquad\dot{\eta}(t)=a(t)^{-1}.
\end{equation}
Differentiating \eqref{BDconformal} with respect to cosmic time and using \eqref{eq:deriv-conf} we have
\begin{equation}
    \frac{\dd}{\dd t}v_k(\eta(t))=-\frac{i k}{a(t)}v_k(\eta(t)).
\end{equation}
Evaluating again at the same initial time $t_i$ than in \eqref{eq:BD-init1} we have, for all modes, that
\begin{equation}\label{eq:BD-init2}
    \dot{v}_k(t_i)= -\frac{i}{a(t)}\sqrt{\frac{k}{2}}\,e^{-ik\eta(t)}.
\end{equation}
Therefore, from Eqns. \eqref{eq:BD-init1} and \eqref{eq:BD-init2} we have the following real and imaginary initial conditions
\begin{equation}
    \operatorname{Re}\left[v_k(t_i)\right]=\frac{1}{\sqrt{2k}}, \qquad
    \operatorname{Re}\left[\dot{v}_k(t_i)\right]=0, \qquad
    \operatorname{Im}\left[v_k(t_i)\right]=0, \qquad 
    \operatorname{Im}\left[\dot{v}_k(t_i)\right]=-\frac{1}{a}\sqrt{\frac{k}{2}}.
\end{equation}

%%%%%%%%%%%%%%%%%%%%%%%%%%%%%%%%%

%\section{Collapsing time}\label{subsec:time}

%\tcr{Following \cite{Goncalves:2000nz} and the two appendices of \cite{Martin:2019nuw}, we have that for small over-densities ($\delta_k<1$), the time a density perturbation will spend to collapse into a black hole is given by:
%\begin{equation}\label{eq:collapse-time}
%    \Delta t_{\rm coll} = \frac{\pi}{H\delta_k^{3/2}}.
%\end{equation}
%This collapsing time should be considered when computing the mass fraction since, depending on it, the PBH can form or dissipate.
%\begin{figure}
 %    \centering
 %    \begin{subfigure}[b]{0.49\textwidth}
 %        \centering         \includegraphics[width=\textwidth]{Images/ncoll-65.pdf}
  %       \caption{}
  %       \label{fig:ncoll-65}
  %   \end{subfigure}
  %   \hfill
   %  \begin{subfigure}[b]{0.49\textwidth}
   %      \centering         \includegraphics[width=\textwidth]{Images/ncoll70.pdf}
   %      \caption{}
    %     \label{fig:ncoll-70}
    % \end{subfigure}
    %    \caption{Time needed for collapse (translated into a number of e-folds since collapse starts, see \eqref{eq:n-coll}) as a function of the comoving wavenumber $k$. In a), we assume that collapse starts at $N=65$ and in b) at $N=70$. The vertical continuous grey line is at $k_{\text{end}}$ and separates the type I from the type II modes. Vertical grey dashed lines in both plots mark the smallest comoving Jeans length (right) and the comoving Hubble radius (left).}
        \label{fig:ncoll}
%\end{figure}
%Fig.~\ref{fig:ncoll} shows the collapsing time assuming that the collapse starts at two different moments, at $N=65$ in Fig.~\ref{fig:ncoll-65} and $N=70$ in Fig~\ref{fig:ncoll-70}. The cosmic time $\Delta t_{\text{coll}}$ has been translated into a number of e-folds since collapse starts. We have selected just the modes that satisfy the Jeans length argument \eqref{eq:jeansargument}, and we have also considered that all modes start to collapse at the same time. We can observe that for the modes with higher $k$ (and therefore higher $\delta_k$) the collapse needs fewer e-folds to occur, see \eqref{eq:collapse-time}. Thus, it is expected that if PBH forms, it will be easier for the modes with higher $k$ and also for modes collapsing at a later time since, in general, modes seem to need less $\Delta N_{\text{coll}}$ to collapse for Fig~\ref{fig:ncoll-70}. The procedure to translate $\Delta t_{\text{coll}}$ into $\Delta N_{\text{coll}}$ is made as follows. Assume that collapse starts at $t=t_{\text{start}}$, then using the definition of number of e-folds $\dd N=\dd\ln{a}$ we have that
%\begin{equation}\label{eq:n-coll}
%    \Delta N_{\text{coll}}=\ln{\left(\frac{a(t_{\text{start}}+\Delta t_{\text{coll}})}{a(t_{\text{start}})}\right)}.
%\end{equation}
%In this manner, we start to count the number of e-folds from the moment the collapse starts. To compute $\Delta N_{\text{coll}}$ in Fig.~\ref{fig:ncoll} we just substituted $t_{\text{start}}$ by the corresponding time to $N=65$ and $N=70$.}

%%%%%%%%%%%%%%%%%%%%%%%%%%%%%%%%%%%

\acknowledgments
Daniel del-Corral is grateful for the support of grant UI/BD/151491/2021 from the Portuguese Agency Funda\c{c}\~ao para a Ci\^encia e a Tecnologia. This research was funded by Funda\c{c}\~ao para a Ci\^encia e a Tecnologia grant number UIDB/MAT/00212/2020   and COST action 23130. NSF grant PHY-2014075 supports the work of P.G. 
P.G. thanks Profs. Masahide Yamaguchi and Teruaki Suyama for their gracious hospitality at the Tokyo Institute of Technology, Tokyo, Japan, where parts of the research in this article took place. 
KSK acknowledges financial support from JSPS and KAKENHI Grant-in-Aid for Scientific Research No. JP20F20320 and No. JP21H00069. KSK would like to thank The Royal Society for the support in the name of Newton International Fellowship. KSK would like to thank Prof. David Wands for useful discussions. We thank the anonymous referees for useful comments. 

%\paragraph{Note added.} This is also a good position for notes added
%after the paper has been written.


\bibliographystyle{utphys.bst}
\bibliography{PrePBH.bib}

%\begin{thebibliography}{99}

%\bibitem{a}
%Author, \emph{Title}, \emph{J. Abbrev.} {\bf vol} (year) pg.

%\bibitem{b}
%Author, \emph{Title},
%arxiv:1234.5678.

%\bibitem{c}
%Author, \emph{Title},
%Publisher (year).

% Please avoid comments such as "For a review'', "For some examples",
% "and references therein" or move them in the text. In general,
% please leave only references in the bibliography and move all
% accessory text in footnotes.

% Also, please have only one work for each \bibitem.

%\end{thebibliography}
\end{document}